# Dilatancy in dense suspensions of model hard-sphere-like colloids under shear and extensional flow

Ricardo J. E. Andrade[1,2], Alan R. Jacob[2], Francisco J. Galindo-Rosales[3], Laura Campo-Deaño[3], Qian Huang[4], Ole Hassager[4], George Petekidis[2]

Present Address: [1]MackGraphe – Graphene and Nanomaterials Research Center
Mackenzie Presbyterian University, São Paulo, Brasil

[2]Institute of Electronic Structure & Laser (IESL)
Foundation for Research & Technology-Hellas (FORTH), Heraklion, Crete, GREECE

[3]Centro de Estudos de Fenómenos de Transporte (CEFT)
Faculdade de Engenharia da Universidade do Porto, Porto, Portugal

[4]Department of Chemical and Biochemical Engineering,
Technical University of Denmark, Lyngby, Denmark

Abstract

Dense suspensions of model hard-sphere-like colloids, with different particle sizes, are examined experimentally in the glass state, under shear and extensional rheology. Under steady shear flow we detect Discontinuous Shear Thickening (DST) above a critical shear rate. Start-up shear experiments show stress overshoots in the vicinity of the onset of DST related with a change in microscopic morphology, as the sample shows dilatancy effects. The analysis of the normal stress ($\sigma_N$) together with direct sample observation by high speed camera, indicates the appearance of positive $\sigma_N$ and dilation behavior at the shear thickening onset. Dilatancy effects are detected also under extensional flow. The latter was studied through capillary breakup and filament stretching experimental setups, where liquid-like response is seen for strain rate lower than a critical strain rate and solid like-behavior for higher strain rates. Monitoring the filament thinning processes under different conditions (volume fractions and strain rates) we have created a state diagram where all responses of a hard-sphere suspension (Newtonian, shear thinning, shear thickening, dilatant) are shown. We finally compare the shear thickening response of these hard-sphere-like suspensions and glasses in shear with that in extensional flow.

## 1. INTRODUCTION

Concentrated colloidal suspensions often show an increase in shear viscosity, beyond a critical shear rate, a phenomenon that is known as shear-thickening. Such behavior has been observed in a wide variety of suspensions [1,2], and can often cause damage in industrial processes by breaking pumps and damaging valuable processing equipment. Understanding and controlling shear-thickening behavior is important, not only in the design of industrial processes, but also on a wide range of innovative applications such as bullet proof body armor [3], impact protective sandwich structures [4], damping devices [5,6] and driving fluids for enhanced oil recovery [7]. Despite the importance of such phenomenon, the mechanism behind the shear thickening of concentrated is still under debate.

Different mechanisms have been proposed to explain aspects of shear-thickening over the years. The first attempt to understand the relevant microstructural rearrangements came from Hoffman [8–10] who observed that polymer particles formed layers and moved across each other at low shear rates, causing shear thinning, whereas at higher shear rates a disordered regime was related with shear-thickening. However, later it was shown that shear-thickening can occur without necessarily an order-disorder transition taking place [11,12] suggesting that such microstructural reorganization is not a prerequisite.

Brady and Bossis [13] introduced the hydrocluster mechanism. They suggested the formation of hydroclusters, temporary stress bearing aggregates of particles, that form as a result of short-range hydrodynamic forces and overcome the repulsive forces between particles during shear. Experimentally, this mechanism has been probed by rheo-optical measurements [14–16], neutron scattering [17–20] and by confocal microscopy [21]. This mechanism is being demonstrated to be responsible for the Continuous Shear Thickening (CST), where viscosity increases gradually, [13,22,23] in Brownian suspensions of moderate volume fractions [21,24]. However, this mechanism alone does not seem to predict quantitatively the Discontinuous Shear Thickening observed experimentally, which is related to the abrupt, step rise of viscosity with increasing shear rate.

More recently, frictional contacts have been introduced as a key underlying mechanism in DST of non-Colloidal suspensions [25–28] that is expressed in extreme cases by dilation [29–32]. These studies suggest that when the applied shear stress overcomes the lubrication film, contact between particles takes place enabling local friction between particle surfaces to develop. These frictional contacts produce shear stress proportional to normal stress, leading to a dramatic increase in the measured shear stress, which coincides with shear-thickening. Moreover, at sufficiently high volume fractions, particle crowding and jamming [33–36] promote frictional contacts and then dilation forces that push against boundaries, usually at a liquid-air interface, are introduced. This mechanism was able to predict the DST, however most of the experimental data have been performed on non-Brownian suspensions.

Theoretical models also suggest that while DST in jammed states results in flow cessation, for concentrated suspension below the glass transition DST is related with shear rate hysteresis where shear thickened system still flows under shear [37]. Recent experiment focused on separating the contributions from hydrodynamic interactions and contact forces using shear-reversal techniques [38] and compared viscosity behavior to friction-based models [39], or used topographically rough particles to evaluate friction effects on shear-thickening [40]. The signature difference between friction and lubrication forces is the stress anisotropy originating from the shear-induced microstructure involved in these two types of mechanisms [40,41]. This anisotropy is reflected in the first normal stress difference $N_1$ as it was shown by simulations based on hydrodynamics interactions showing shear-induced distortions and short ranged lubrications forces lead to $N_1 < 0$ [22,24,42,43]. It was also demonstrated that the sign of $N_1$ does not change with repulsive interactions or elastic deformations [36,41,44]. On the other hand, positive normal stress have been related to dilatancy, as a feature of dense and frictional granular materials [31,45]. It is important to refer that in these works, dilatancy is a direct physical consequence of outward-pushing force (positive normal stress), rather than first normal stress difference $N_1$ [46,47]. At high volume fractions, contact forces lead to dilation ($N_1 > 0$), due to the anisotropy of the force chain network [48], as it occurs in granular systems [49]. Due to the experimental difficulty studies providing accurate $N_1$ values in shear thickening suspensions are limited, with most experiments showing $N_1 < 0$ [50–53], with the exception of experiments with rough particles [33]. Recently, it was demonstrated that at moderate volume fractions, negative contributions of $N_1$ from lubrication forces can hide positive frictional contributions, but at significant high volume fractions and stresses, frictional interactions become dominant causing positive $N_1$ [41,54].

As stated above, measuring the first normal stress difference can provide valuable information of the micro-macro relationship; moreover, by subjecting a shear-thickening suspension to varying types of flow deformations it is possible to access the dynamics and physics of the suspension. In this context, extensional flows are also very important to colloidal suspensions. Although concentrated colloidal suspensions have attracted considerable attention in shear flows, the studies in extensional flows are still limited.

Extensional flow usually enhances the non-linear behavior, and exhibit dynamics and structural changes different from the shear flow [47]. Extensional flows are of great interest because many industrially relevant processes, such as agro-chemical spraying, enhanced oil recovery, coating flows, fiber spinning, and food processing, are dominated by the fluids' extensional properties [7,55,56]. Bischoff and White *et al.* [57] investigated the extensional rheology of corn starch in water suspensions using a home-built Filament Stretching Extensional Rheometer (FiSER) and Capillary Breakup Extensional Rheometer (CaBER), to study the mechanism of strain hardening. A strong extensional hardening behavior, which means increase in the extensional viscosity with strain rate, was observed beyond a critical extension rate, which was related to the aggregation of particle clusters. Chellamuth *et al.* [58] studied the extensional rheology of dispersion of fumed silica particles suspended in polypropylene glycol with low molecular weight using a FiSER set-up combined with light scattering measurements to elucidate the microstructure evolution during flow. Beyond a critical extensional rate, a dramatic increase in strain-hardening effect was observed, similar to the shear thickening transition observed under shear, however with a higher magnitude and at reduced critical deformation rates. Light scattering measurements showed that the extensional hardening was due to the alignment of nanoparticles and the formation of large aggregates in the flow direction. These were the first direct observations of hydrodynamically induced clustering in extensional flow. Smith *et al.* [59] analyzed the flow behavior of concentrated suspensions of nearly hard-spheres colloids under extensional flow with a single particle size at different volume fraction. They showed that above a critical extension rate, the suspension jams and dilates, by expanding its volume, exhibiting dramatic granulation as demonstrated by a macroscopic lumpiness and rough surface appearance, followed by a solid-like break up behavior. Besides the above studies until now, there is a lack of information that relates particle size in concentrated colloidal hard sphere suspensions with the strain hardening and consequent dilatancy behavior in extensional flow, and its direct comparison with thickening under shear flow.

In this work, we present an analysis of the influence of the particle size in the rheological behavior of concentrated colloidal dispersions under shear and extensional flow. To access the dynamics and physics of the suspension, a complete rheological characterization is necessary to understand the relation between micro- and macro-scale of these systems. More specifically, we study the rheological response of model nearly hard sphere colloids of three different sizes all at the colloidal scale (i.e. where Brownian motion is dominant at rest), under shear and extensional flow. Depending on the size and rate of deformation, different behaviors were observed. The combination of the information provided by both types of flow, allows a comprehensive characterization of the system, providing information of microstructure and dynamics related with shear-thickening. Our work here complements recent studies on shear thickening by addressing how ideas of dilatancy, which were mostly developed for non-colloidal particles, apply to colloidal suspensions, and presents an example where an interplay between dilatancy and (presumably) lubrication-induced negative normal stresses takes place in shear experiments. We further relate such shear thickening response with corresponding findings in extensional rheometry.

## 2. EXPERIMENTAL SECTION
### A. Samples

Colloidal glasses containing nearly hard sphere (HS) particles with sizes approximately R ≈ 689, 405, and 137 nm and polydispersity ranging from 7% to 10%, were prepared. Particles are made of poly-methylmethacrylate (PMMA) sterically stabilized by a thin grafted layer (≈ 10nm) of poly-12-hydroxystearic acid (PHSA) chains inducing steric repulsion that stabilizes them against van der Waals attractions creating an almost ideal hard sphere interaction. PMMA particles (density ≈ 1180 kg/m$^3$) were dispersed in squalene (viscosity = 0.015 Pas), a non-volatile solvent with refractive index of 1.494 , very similar PMMA, therefore minimizing any residual van der Waals attractions [60]. The surface tension and density of squalene is 29.2 mN/m and 848 kg/m$^3$ at 25 ºC, respectively. The largest particles (R ≈ 689 nm) were used in the glassy ($\varphi \approx 0.60$) and in supercooled regime ($\varphi \approx 0.56$) while the smallest ones were used only for measurements the glassy regime ($\varphi \approx 0.60$). These volume fractions were chosen with the aim to investigate the shear thickening-behavior in the glassy regime where the onset is detected at lower stresses and shear rates [2,16,61]. The bare Brownian time defined as $t_B = 6\pi\eta_s R^3/k_B T$ was used in the non-dimensional shear rate, $Pe = \dot{\gamma} t_B$ that quantifies the interplay between Brownian motion and shear, and similarly in the non-dimensional extension rate, $Pe_{ext} = \dot{\varepsilon} t_B$. The Brownian times for the particles with radii 689nm, 405nm and 137nm are 22.87 s, 4.645 s, and 0.0179 s, respectively.

### B. Shear rheology

A cone and plate geometry attached to an Anton Paar MCR 501 rheometer was used for the shear experiments. More specifically we used a glass cone with diameter 40 mm, angle = 1.967° and truncation gap = 176 μm. The temperature was fixed at 20 °C with a Peltier based thermostat. Dynamic frequency sweep experiments together with flow curves provide a complete characterization of the rheological behavior of the samples. Start-up shear measurements at a fixed shear rate were performed to follow the transient response. All tests were conducted after shear rejuvenation at $\dot{\gamma} = 0.5$ s$^{-1}$ for 300 s followed by a waiting time of 200 s. The final value of the viscosity determined during steady state for each start-up shear rate test is used to construct the steady state viscosity flow curve (stress vs. shear rate). The first normal stress difference $N_1 = \sigma_{xx} - \sigma_{yy}$, where $\sigma_{xx}$ is the stress acting normal to the shear plane and $\sigma_{yy}$ is the stress acting normal to the shear gradient plane, extracted from the axial force measured by the transducer, $2F_y = N_1 \pi R^2$ [46,47]. However, the above hold under certain conditions including that, in spherical coordinates, the velocity in the radial (3) and θ (2) directions is zero (also angle of cone is small, spherical liquid boundary). Hence while these conditions can be assumed true in the shear thinning regime, they break down when the sample dilates. Thus to avoid any confusion we clarify $N_1$ documented here can be considered as the first normal stress difference up to the point where the conditions hold, i.e. before dilatancy sets in, while when DST and dilatancy are present it represents the normal (axial) force, $F_N$, per unit area (or normal stress, $\sigma_N$).

A high-speed Basler camera was used to visualize the edge effects during shear experiments in order to relate shear thickening with dilatancy and edge fracture effects. Images were recorded every 100 ms for a specific amount time depending on the duration of the shear experiment.

## C. Extensional rheology

The rheological characterization of these concentrated colloidal suspensions under extensional flow conditions were performed at 23 ºC by means of both extensional rheometers available in the market for characterizing the extensional properties of complex liquids [56], i.e. the Capillary Breakup Extensional Rheometer (Haake CaBER-1) and the Filament Stretching Rheometer (FSR) (VADER 1000 by Rheofilament).

CaBER imposes an extensional step-strain and the filament subsequently thins exclusively under the influence of capillary forces. The extensional strain rate is the result of a balance between the interfacial tension and extensional properties of the liquid. In the CaBER device, a laser micrometer uniquely measures the time evolution of the diameter of the thinning filament at the mid plane, $D_{mid}(t)$, which ideally coincides with the minimum diameter, $D_{min}(t,)$ and from which the strain rate at a specific time, $\dot{\varepsilon} = \frac{d\varepsilon}{dt} = -\frac{2}{D_{min}(t)}\frac{dD_{min}(t)}{dt}$, and the apparent extensional viscosity, $\eta_{Eapp} = -\frac{\sigma}{(dD_{min}(t))/dt}$, can be calculated [62], using the surface tension, $\sigma$. However, as it has been already reported before [63–66] the shape of the filament depends on the interplay between the rheological response of the complex fluids and the test conditions. Therefore, the minimum diameter can be out of the mid-plane. It is then required to use a high-speed camera to monitor the filament thinning dynamics after a rapid extensional deformation. A digital post-processing image analysis in Matlab allows in this way to determine $D_{min}(t)$. The experimental setup was similar to the one in Campo-Deaño and Clasen [64], in the sense that the filament thinning processes were recorded by means of a high-speed video camera (Photron FASTCAM Mini UX100) recording up to 5000 fps (1280 × 1024 square pixel) with an average image resolution of ≈ 5 μm per pixel, in substitution of the laser detector. It is important to highlight that no SRM (Slow Retraction Method) was imposed in the CaBER experiments. The configuration used in our work, however, did not allow recording the filament dynamics simultaneously with the laser-based measurement. Thus, we performed independent experiments determining $D_{mid}$ both with the laser and with camera assisted configurations. In Figure 1a we show that the mid plane diameter and the minimum diameter do not actually coincide when the sample thickens under extensional flow, leading to inaccurate measurements when the determination of $D_{mid}$ is performed just with the laser of assisted setup. For this reason, hereafter only $D_{min}$ measured with the camera will be presented in the following figures.

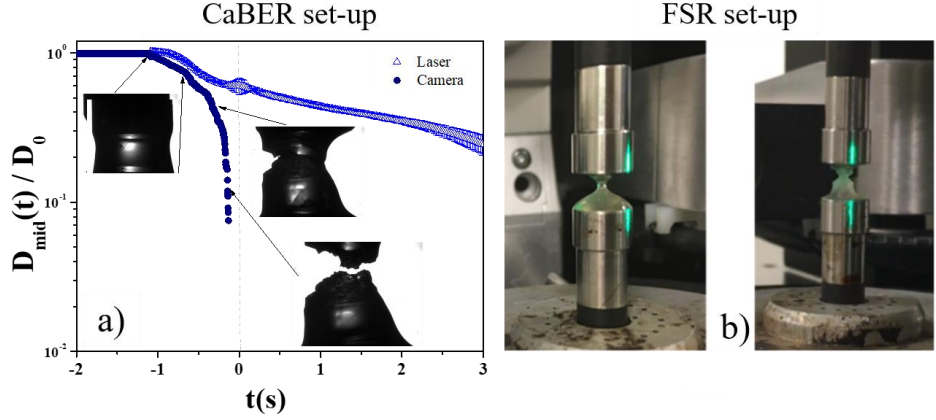

Figure 1. (a) Time evolution of the stretched filament diameter in a CaBER experiment. Comparison between $D_{mid}(t)$ obtained with laser (dark blue symbols) and $D_{min}(t)$ obtained from the analysis of the images recorded with a high-speed camera (light blue symbols) as a function of time during a capillary breakup extensional rheology measurement (Sample with HS particles of R ≈ 405 nm at strain rate $\dot{\varepsilon}_0$ = 1 s$^{-1}$). (b) Images from FiSER set-up with a HS glass of particles with R ≈ 689 nm during stretching at a low strain rate, 0.1 s$^{-1}$, showing a liquid-like behavior (left) and a high strain rate, 1.0 s$^{-1}$, showing solid-like behavior (right).

It is known that low values for the initial aspect ratio ($\Lambda_0 = \frac{2h_0}{D_0}$), where $h_0$ is the initial distance between the end-plates and $D_0$ the initial sample diameter, promote inhomogeneity during extensional flow, as the radial velocity fluctuations dominate fluid motion in the filament generating shear flow near the endplates. During the early stages of the experiment this is proportional to $\frac{\dot{\varepsilon}}{\Lambda_0^2}$ [57,67]. Nevertheless, due to the low surface tension of the sample, we were forced to impose $\Lambda_0 = 0.75$, since the initial gap must be smaller than the capillary length ($h_0 \leq \sqrt{\sigma/\rho g}$), where $\sigma$ is the surface tension, $\rho$ is the density and $g$ is the gravitational acceleration, in order to hold the sample between the plates and run the experiments [63].

Thus, the time evolution of $D_{min}$ obtained from the CaBER has two zones. The first one, with negative times (from -$t_s$ to 0, where $t_s$ is the streak time), corresponding to the stretching zone, in which the top endplate has to reach to the final height ($h_f$) and during which the fluid flow is not purely elongational, but with an important shear contribution. The strain rate imposed to the sample is therefore defined as $\dot{\varepsilon}_0 = \frac{h_f - h_0}{h_0 t_s}$ (see Table I). The second zone, at positive times (t>0), is where the capillary forces thinning the filament are counterbalanced by the microstructural forces of the sample until it breaks up. In this study, the total time to break the filament ($t_{br}$) is considered from the moment in which the top plate starts to move until the sample is not bridging the plates anymore.

Table I. Experimental conditions imposed in the linear stretching profile imposed in the CaBER experiments.

| $h_0$ [mm] | $h_f$ [mm] | $t_s$ [s] | $\dot{\varepsilon}_0$ [s$^{-1}$] |
|---|---|---|---|
| 1.5 | 4.450 | 19.66 | 0.10 |
| | | 7.866 | 0.25 |
| | | 3.933 | 0.50 |
| | | 1.966 | 1.00 |
| | | 0.988 | 1.99 |
| | | 0.655 | 3.00 |
| | | 0.393 | 5.00 |
| | | 0.039 | 50.4 |
| | 19.64 | 0.096 | 126 |
| | | 0.020 | 605 |

The filament stretching rheometer (FSR) (Figure 1b) imposes a velocity on the upper plate in order to induce uniaxial extensional flow. The temporal evolution of the tensile force, *F(t),* exerted by the fluid column on the bottom plate and on the filament radius at the axial mid-plane, *D$_{min}$(t)*, are measured by a load cell and a laser based micrometer, respectively, and are used to compute the transient extensional viscosity [68].

The time evolution of the uniaxial extensional stress was measured in the FSR for each sample. The samples were loaded between the two plates with a diameter of 9 mm. The height of the samples was around 1.75 mm ($\Lambda_0 = 0.389$). This initial aspect ratio introduces an important shear component in the flow at early stages of the stretching, as it was noted before. Unlike the CaBER, the FSR allows us to perform a two-step stretching protocol. According to White et al. [57], this protocol is similar to applying a pre-shear step in shear experiments on a rotational rheometer in order to erase pre-shear history (rejuvenation) of the sample. Here, an initial stretch at extension rate well below the onset of extensional thickening is applied, followed by a second stretch much faster until the filament breaks up. The first step thus aims to minimize the shear effects on the second step, which is the actual experimental measurement.

The Hencky strain and the strain rate are calculated for the FSR set-up as $\varepsilon(t) = -2\ln(D_{mid}(t)/D_0)$ and $\dot{\varepsilon} = \frac{d\varepsilon}{dt}$. To ensure a constant strain rate, the mid-diameter is required to decrease exponentially with time during stretching. A close loop control scheme which consists of a feed-forward and a feed-back loop [69,70] is employed in the FSR to ensure accurate constant stretch rate. The close loop control scheme is limited by a maximum stretch rate of 5 s$^{-1}$. For the rates faster than this, an open loop control scheme that switches off the feedback contribution has to be used. In the latter case, the feed-forward control parameters, which define the velocity of the top plate, were obtained by trial-and-error until the mid-plane diameter decreases exponentially with time [71].

Finally, it is important to refer that in both extensional rheometers, CaBER and FSR, the fluid sample was always pinned to the edges of the plates.

## 3. RESULTS AND DISCUSSION

### Shear rheology

*Linear viscoelasticity - Flow curves*

The linear viscoelastic response of colloidal suspensions with similar volume fractions and different particle sizes is presented in Figure 2. To allow a direct comparison of different systems the elastic and viscous moduli, $G'$ and $G''$, are normalized by thermal energy density, and the frequency is multiplied by the Brownian time, $Pe_\omega = \omega t_B = \omega 6\pi\eta_s R^3/k_B T$. The two smaller particle size samples with φ ≈ 0.6 (in the glass regime) show frequency dependent moduli with a solid-like behavior where G' dominates over $G''$ at low *Pe*, indicative of the kinetically trapped particles within a cage formed by their neighbors. At high frequencies, $G'$ and $G''$ approach each other at a characteristic crossover frequency and exhibit a liquid-like behavior with $G'' > G'$ at higher frequencies, a response related with local in-cage particle diffusion. For the larger particle size (R ≈ 689 nm) samples the high frequency regime liquid-like response due to in-cage motion is seen better (as the $Pe_\omega$ extends to higher values) both in the glass sample (φ ≈ 0.6) and the supercooled liquid (φ ≈ 0.56).

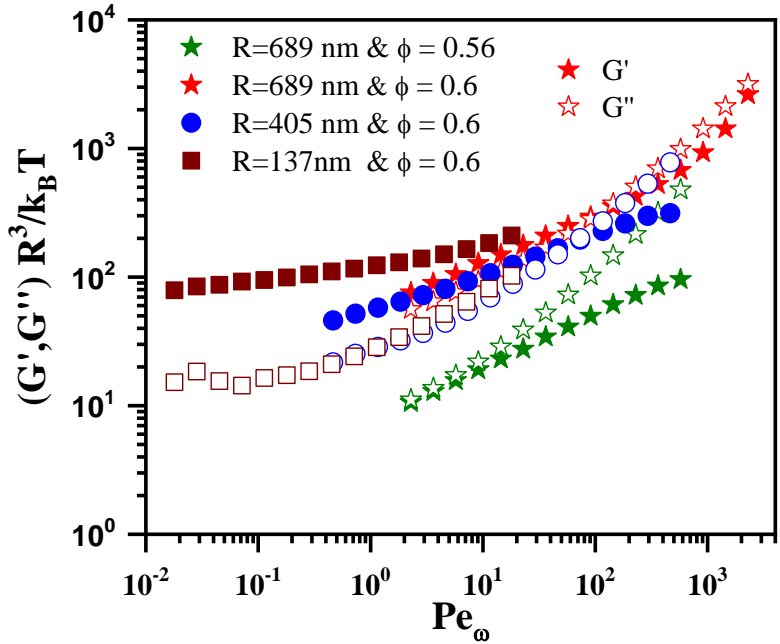

Figure 2: Dynamic frequency sweeps for different particle sizes as indicated in squalene. Solid symbols represent the elastic moduli, *G'*, and open symbols the viscous moduli, *G''*, both scaled by $\frac{k_B T}{R^3}$ versus scale frequency $Pe_\omega$.

Figure 3a shows flow curves for the same systems where the viscosity was measured as function of shear rate at steady state following step-rate experiments as explained above. With the exemption of the smaller particle (R ≈ 137 nm) all samples show a sharp increase of the viscosity at a critical shear rate, indicating a Discontinuous Shear Thickening (DST) behavior. The critical shear rate for the onset of DST decreases with increasing particle size as expected [61,72]. With the aim to scale out the effect of particle size, the data are also plotted versus the dimensionless shear rate, i.e. *Pe* in Figure 3b. More specifically we detect the onset shear rate of DST at around 0.6 s$^{-1}$ and 3 s$^{-1}$ for particles with R ≈ 689 nm and R ≈ 405 nm, respectively, which correspond to similar values of *Pe* (≈ 13.72 and 13.94, respectively) for the glassy samples of the two larger particles at φ ≈ 0.6. This finding suggests that at the same volume fraction a critical Pe needs to be exceeded in order for DST to take place. Note that the sample with the smaller PMMA particles (R ≈ 137nm), at the same volume fraction (φ ≈ 0.6), does not show any DST as this critical Pe (of the order of 10) has not been reached (see figure 3b). This is similar to the what is observed in Figure 2, where for the smaller particle size sample data due not reach the high frequency regime. The critical shear rate for the onset of DST is increasing with decreasing volume fraction as demonstrated by the measurement with the largest particles at φ ≈ 0.56, just below the glass transition.

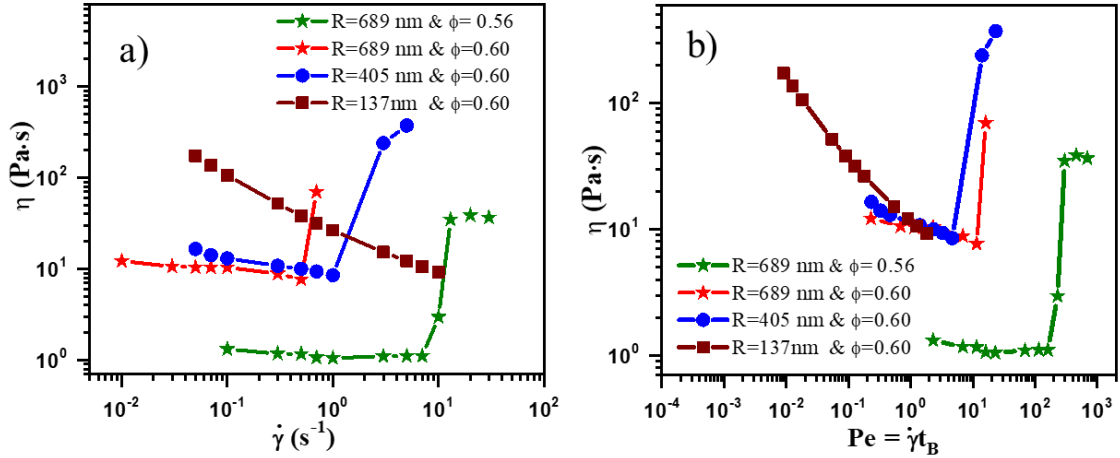

Figure 3: Flow curves for all samples measured with different particle sizes and volume fractions as indicated (a) steady state stress plotted as a function of shear rate and (b) plotted as a function of dimensionless Pe.

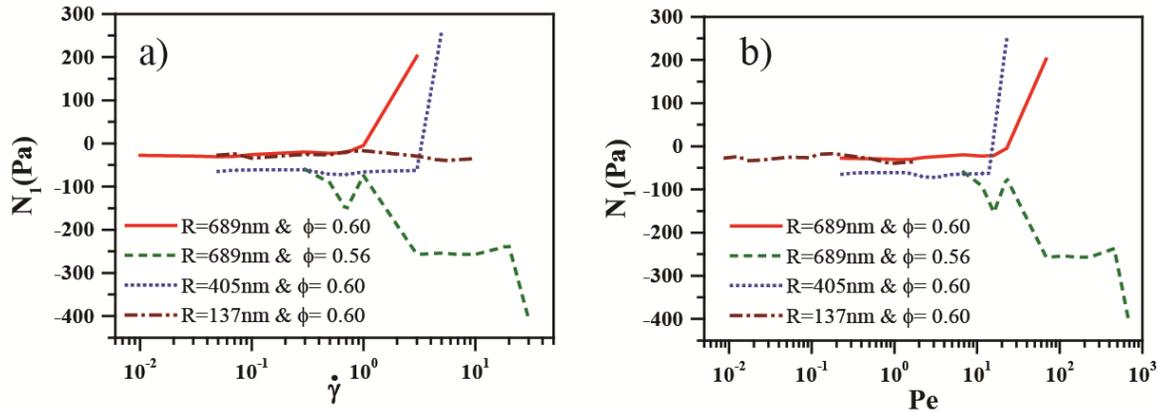

Figure 4: (a) First normal stress difference, $N_1$, as a function of shear rate for all samples measured as indicated. (b) the same data as in (a) plotted versus the dimensionless shear rate, Pe.

*Normal stresses*

The first normal stress difference $N_1$ measured together with the shear stress shown in the flow curves of Figure 3, is plotted in Figure 4a and b, as a function of shear rate and Pe respectively, similarly with the representation of shear stress in figure 3. For all samples up to the critical shear rate where shear-thickening phenomena set-in, we observe $N_1 \approx 0$, suggesting relatively weak structural changes, as it was seen in Figure 3 and discussed above. For the two larger particles (R ≈ 689 nm and R ≈ 405 nm) at the higher volume fraction (φ ≈ 0.6) measured, $N_1$ increases suddenly around the critical shear rate where shear thickening is observed, following the behavior of the shear stress (Figure 3). For the larger particles (R ≈ 689 nm) at lowest volume fraction though (φ ≈ 0.56), $N_1$ starts acquiring negative values at rates much lower (around 1 $s^{-1}$) that the onset of shear thickening (10 $s^{-1}$) and keeps decreasing significantly well in the shear thickening regime.

*Transient response (start-up shear)*

Figure 5 presents the measured stress versus strain during start-up shear tests at different shear rates for all three samples. For the samples presented in Figure 3a and 3b we can observe a pronounced stress overshoot at high shear rates. In general, in colloidal suspensions and glasses the stress overshoot is related to structural deformation of the first neighbor cage as stress is stored during an initial deformation stage and then a release upon cage breakage and flow [73,74]. This takes place at strains that can be identified as the yield strain which for HS glasses is of the order of 10%, with a weak non-monotonic volume fraction dependence and a decrease towards zero at random close packing. In the experiments however shown in Figure 5 the stress overshoot is detected at much larger strains (of the order of 100%) indicating that they are of a different origin and not related with yielding but rather with larger length-scale rearrangements. As this take place only for systems and shear rates only where shear thickening is observed (Figure 3 and 4) they must reflect the onset of such phenomena during start-up shear. Moreover, they appear rather complex with second peak observed after the first. Note that the second peak seems to be related with flow instabilities, as it will be discussed below in the *visual observation* section. The onset shear rate for the overshoot observation is similar with the critical shear rate where the shear viscosity exhibits a sharp increase, i.e. the onset of DST (Figure 3). Moreover, the sample with the smallest particles (Figure 5c) does not exhibit any such stress overshoot, in agreement with observations in Figure 2b where shear thickening is absent for this sample in the experimental shear rate window.

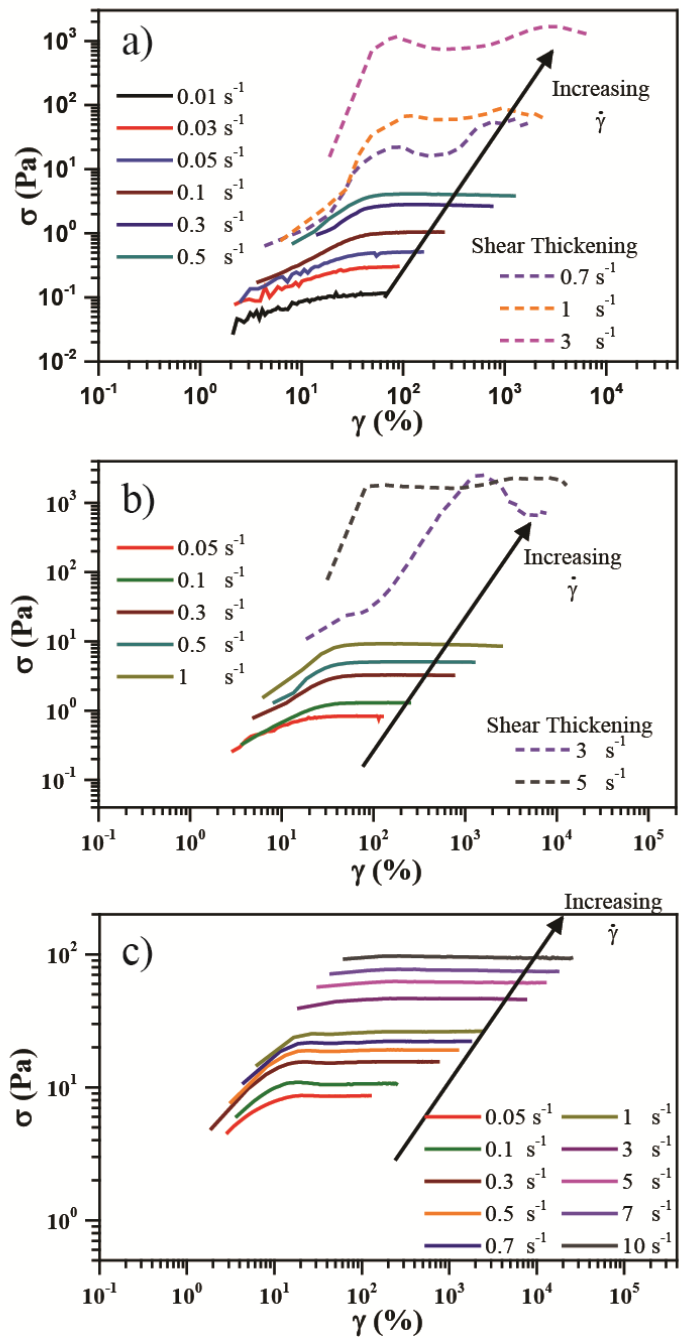

Figure 5: Step-rate experiments (stress versus strain) at different shear rates for all samples with different particles sizes; Large particles R ≈ 689 nm (a), intermediate size particles, and R ≈ 405 nm (b) and small particles, and R ≈ 137 nm (c). All samples were measured at volume fraction of φ ≈ 0.60.

The normal stress $\sigma_N$ is shown in Figure 6 during transient start-up shear experiments for the sample with the two larger particles at φ ≈ 0.6 (corresponding to stress data shown in Figure 5a and 5b). For shear rates below the onset of DST, $\sigma_N$ remains constant and nearly zero. As the shear rate is increased and approaches the vicinity of shear thickening, however, we observe the development of a negative $\sigma_N$ at high strain values (Figure 4), with a minimum near the second stress overshoot and a subsequent increase towards zero at higher strains. This behavior observed for $\dot{\gamma}$ = 0.7 and 1 s$^{-1}$ for particles with R ≈ 689 nm and $\dot{\gamma}$ = 3 s$^{-1}$ for particles with R ≈ 405 nm. The negative $\sigma_N$ values indicate that contributions from lubrications forces become dominant as detected in previous studies [50,51]. Therefore, under such conditions particles collisions may lead to aggregates as result of short-range hydrodynamic forces, overcoming due to the shear flow the repulsive forces between particles and consequently the lubrication film.

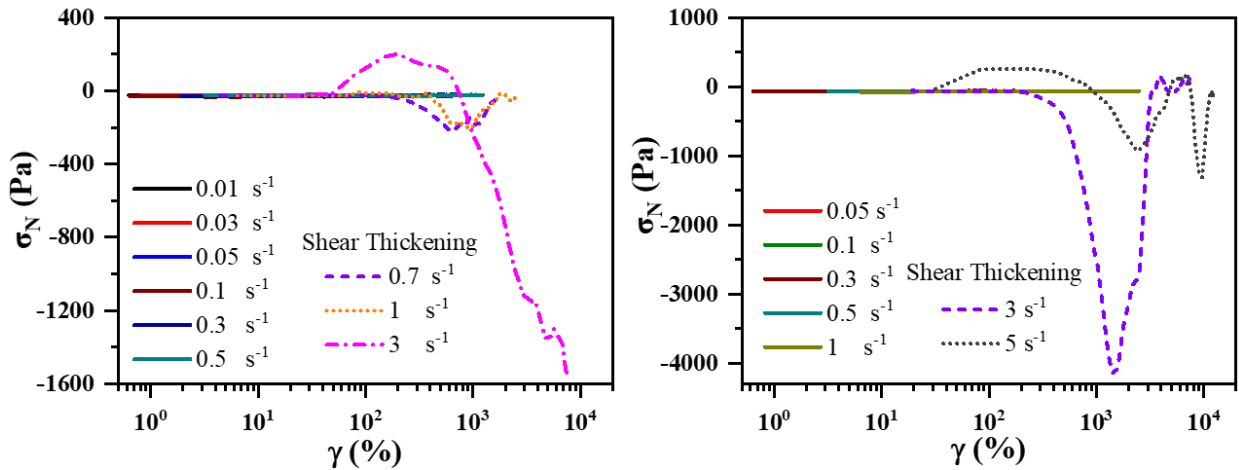

Figure 6: Normal stress $\sigma_N = \frac{2F_N}{\pi R^2}$ during transient start-up shear experiment versus strain, γ, for particles with R ≈ 689nm (left) and R ≈ 405nm (right). All samples were measured at volume fraction of φ ≈ 0.60.

At the highest shear rates measured however, for both particles shown in Figure 6, we detect first an increase of $\sigma_N$ to positive values in the vicinity of the first stress overshoot and subsequently a drop to negative values around strains where the second stress overshoot is detected, as seen at lower shear rates. The transition from $\sigma_N$ > 0 to $\sigma_N$ < 0 with increasing strain, observed during start-up shear at high rates, is indicative of frictional interactions due to the contact between particles [54] around the first stress overshoot, which are somehow relaxed giving way to lubrication forces (responsible for negative $\sigma_N$ values) at even larger strains around and beyond the second stress overshoot. As will be discussed below, the behavior in this regime is mostly due to slip and flow instabilities (see Figure 7 in *visual observation section)*.

*Visual observation*

We complement the rheology data with direct observation of the sample during shear. The edge of the sample was monitored via a video camera during the step rate experiments in order to detect any macroscopic morphological changes related with shear thickening as the strain increases. As at high volume fractions trimming of the excess sample in the gap is a difficult, we leave some excess sample around the cone as seen in the images of Figure 7. Although similar experiments in shear thickening cornstarch suspensions [30] have shown that excess fluid around the geometry may affect the rheological response, as boundary conditions change, we have not observed any clear affect in our measurements; however it should be noted that our experiments were performed in a cone-plate geometry, rather than in a plate-plate with varying gap. Along these lines, we also performed measurements with a cone-partition-plate (CPP) geometry which is often used in polymer melts to suppress edge fracture effects at high shear rates [75]. As shown in the supplementary material the onset of shear thickening both in steady shear measurements as well as in oscillatory shear was not affected by the use of the CPP. Although some details of the curves are different, no significant changes were detected with the use of the CPP.

Figure 7a shows the HS glass with the larger particle size, R ≈ 689 nm, (with φ=0.6) at $\dot{\gamma} = 3 \text{ s}^{-1}$ (or Pe ≈ 68.61), a value which is above the critical value where DST is detected. The evolution of the shear stress and the normal stress $\sigma_N$ are shown during the start-up shear test as a function of strain along with corresponding images of the sample edge. The shear stress is increasing until point B where the first stress overshoot at strains above 100 %, related with DST, is seen as mentioned above. At this point, we detect a slight increase in $\sigma_N$ (with positive values) which continues until point C. Here, we visually observe that the surface of the sample turns opaque and rough, a signature of dilatation of the material. This response at high shear rates can be understood as follows: HS glass starts to flow under the application of constant shear rate after the yield strain (typically of the order of 10% in the glassy regime) is exceeded, as shown in Figure 5 for low rates and demonstrated in previous studies [73,76]. At high shear rates though, where the stress cannot be detected in the linear regime, i.e. at early times and low strains, the sample shear thickens, and a further stress increase and overshoot is observed at strains above 100%. The stress overshoot is then accompanied by a slight increase in $\sigma_N$ which suggests the lubrication film between the particles is overcome by the exerted stress leading to direct contact between the particles. Since the system continues to be under constant shear rate, it expands and dilates in order to allow flow to occur along with some stress/energy release. Beyond point C, a second stress overshoot is detected (see also Figure 5a), the behavior is rather erratic as slip and instabilities set in as seen by the direct visualization of images D and E in Figure 7a. We should also note that after flow cessation the sample exhibits a rather quick healing in very short time span of the order of 1 s as is shown, for the case of the larger particle size in figure 7a.

Similar, but not identical, response was observed in glasses (with $\varphi \approx 0.6$) of the intermediate size particles, R ≈ 405 nm. For the same shear rate as the one shown for the larger size particles, i.e. at $\dot{\gamma} = 3$ s$^{-1}$ (or $Pe = 12.15$) the stress overshoot takes place when higher strains (more than 1000 %) are reached (see Figure 7b). Here the shear thickening related stress overshoot takes place at point F where slip and instabilities develop. In this case $\sigma_N$ stays around zero and, turns into negative values as shear thickening occurs, indicating that hydrodynamic forces are predominant in the vicinity of DST [50,51]. At higher strains (points F to I in Figure 7b), past the shear stress overshoot, the normal force increase rise from negative to around zero possibly indicating a transition to a regime where contact forces are also present. For the smaller particles, R ≈ 405 nm, at higher shear rates, $\dot{\gamma} = 5$ s$^{-1}$ (corresponding to $Pe \approx 20.25$) we observe, similarly with the response in the larger particles shown in figure 7a, a stress overshoot at strains around 100 %, that is accompanied by weak increase of the (positive) normal forces in the regime where the sample exhibits a dilatation (see Figure 7c). Again, as the strain increases further, $\sigma_N$ drops to negative values and exhibits large fluctuations while the stress exhibiting a second overshoot, in a regime where visual observation reveals wall slip and edge instability phenomena. As the suspension approaches very large strain limits, during the transient measurements, edge instabilities and slip also contribute to negative $\sigma_N$, as seen in Figure 6 and 7. In other words, the normal stresses are a function of time, so the sign and magnitude depend on how long the steady shear is performed. On the other hand, the steady state values in Figure 4 do not show a drop on $N_1$ for particle sizes, R ≈ 689 nm and R ≈ 405 nm at $\varphi \approx 0.6$. In these cases, the normal stress values were extracted for smaller strain limits where dilation was observed just before the edge instabilities and slip start becoming prominent and affect the normal stress values.

These observations suggest an interplay of hydrodynamic interactions and frictional contacts takes place at dense colloidal suspensions depending on the size and shear rate applied, and related with macroscopic phenomena such as dilatation, wall slip and edge fracture. Such behavior is important from the rheological point of view, as was also demonstrated recently in [77] for dense non-Brownian suspensions. It is also important to refer this phenomenon did not occur for the smallest particle (R ≈ 137 nm) system mostly due to the limited experimental shear rate window, not reaching the critical *Pe* needed. In addition, we cannot exclude the fact the smaller particle is (proportionally) slightly softer as the stabilization layer is always ~10 nm, and thus larger in proportion to the particle radius for smaller particles. Previous work [78] explicitly shows the stability of both small and large PMMA spheres suspended in squalene with no variation in the bulk viscoelastic modulus observed for as long as 20 hours. Moreover, squalene has a refractive index of (~1.494) which is similar to that of PMMA (1.495). Hence, it can be assumed that van der Waals forces have very little effect on the suspension at rest and during flow as stated for other systems [28].

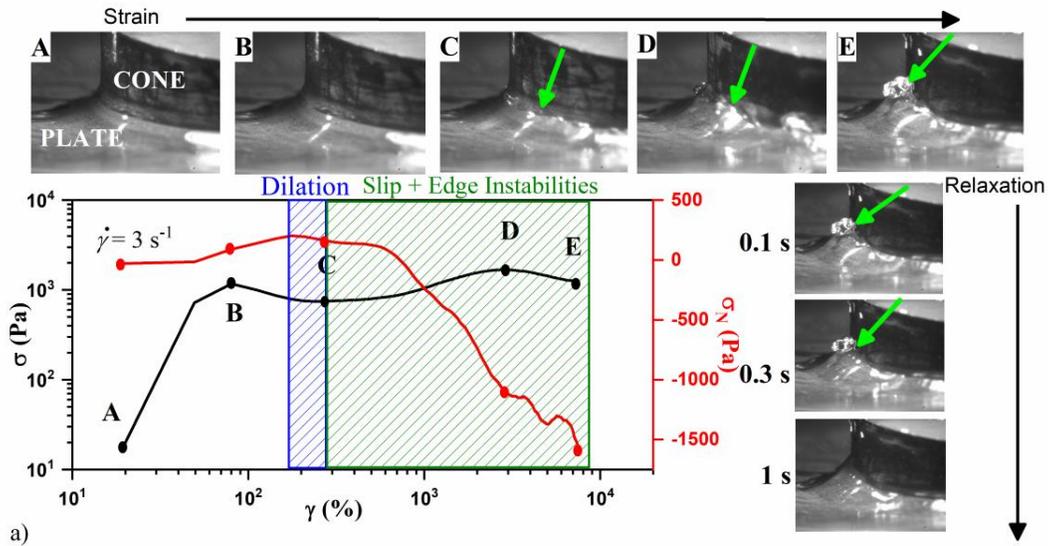

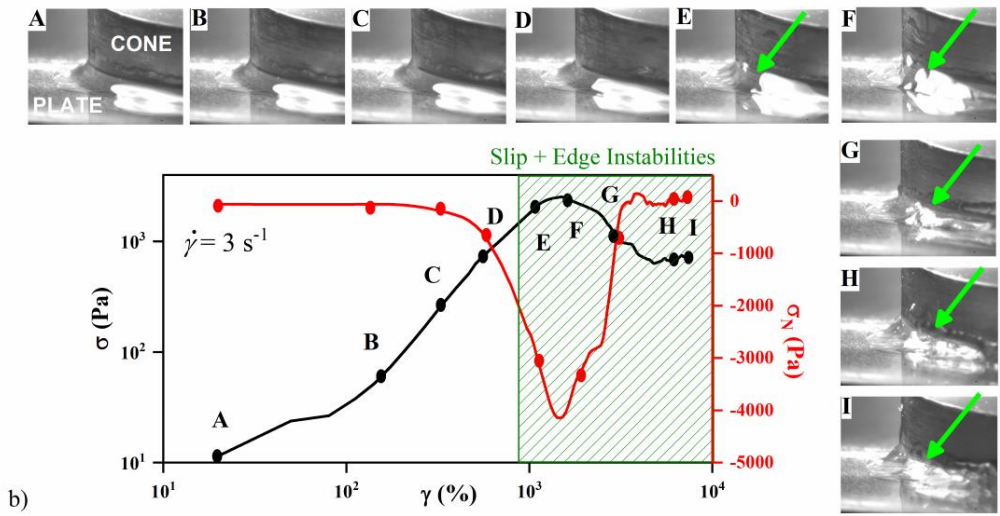

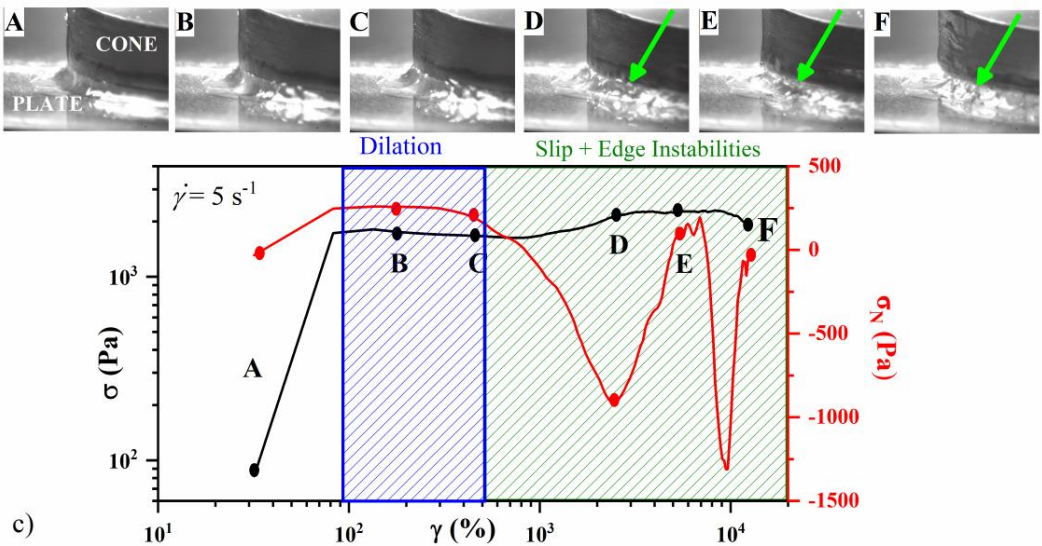

Figure 7: Stages of step-rate experiments monitored by video recorded images following the stress and normal stress $\sigma_N = \frac{2F_N}{\pi R^2}$ as a function of strain for: (a) System with large particles, R ≈ 689nm at constant shear rate $\dot{\gamma} = 3$ s$^{-1}$; (b) Intermediate size particles, R ≈ 405nm, at constant shear rate $\dot{\gamma} = 3$ s$^{-1}$ and (c) Particles R ≈ 405 nm at constant shear rate $\dot{\gamma} = 5$ s$^{-1}$. All samples are at volume fraction of φ ≈ 0.60. In all plots blue regions denote the observation of dilatation, and green regions slip and edge instabilities as indicated by the green arrows. At indicated points within the step rate experiment images of the sample are shown (A to E for (a) and (c) and A to I for (b), while in (a) images of the evolution of the sample after shear is switched off (indicated as relaxation) are also shown at 0.1, 0.3 and 1 s after shear cessation.

### Extensional rheology

We next discuss extensional rheology experiments performed with the same samples, with two different set-ups, CaBER and FSR.

*CaBER experiments*

From the experiments performed with the CaBER set-up, we observe two types of responses as shown in figure 8. For all samples, similarly to shear measurements, there is a critical value of the strain rate, $\dot{\varepsilon}_c$, determined by visual observation and confirmed by Figure 11 as we will discuss, below which the sample is stretched uniformly with a filament created by the end of the initial stretching period. At higher rates, $\dot{\varepsilon}_0 > \dot{\varepsilon}_c$, the samples with the two larger particles exhibit a brittle solid-like response manifested by a break-up with an irregular shape during the stretching period. This behavior is shown in figure 8 (a-d) for all the two large particle size HS glasses, at low (left) and high (right) stretching rates. For the smaller size particle sample the different behavior at the low and high strain rate regime is seen in figure 8e-f; with the high rare response revealing a characteristic pointing shape created during stretching between the two plates [66,79,80].

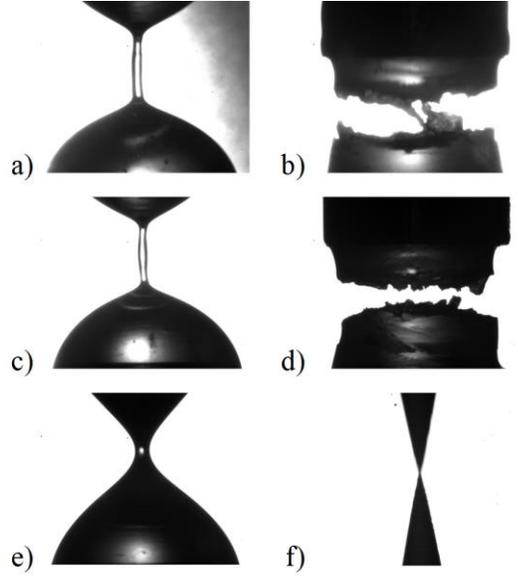

Figure 8. Final filament shape for HS glasses with particle size R ≈ 689 nm (top), R ≈ 405 nm (middle) and R ≈ 137 nm (bottom) for low ($\dot{\varepsilon}_0 < \dot{\varepsilon}_c$, left) and high ($\dot{\varepsilon}_0 > \dot{\varepsilon}_c$, right) shear rates. All samples are at φ ≈ 0.60. These characteristic filament shapes were obtained with the CaBER set-up.

Similarly, with shear experiments below a critical shear (or strain) rate $\dot{\varepsilon}_c$, the surface of the sample remains smooth and the sample is stretched uniformly and symmetrically like a liquid (see Figure 8e and 9c). However, for strain rates higher than $\dot{\varepsilon}_c$ the sample exhibit a solid-like response, losing axial symmetry as they are stretched and exhibiting an irregular shape of the filament as shown in Figure 9a and 9b. In the case of solid-like brittle response the surface texture becomes rough, exhibiting granulation as seen before [59], indicative of microstructural changes in the bulk of the sample. For $\dot{\varepsilon}_0 < \dot{\varepsilon}_c$, samples remain bridging the top and bottom endplates at the end of the stretching zone. Once stretching is completed and the plates are kept constant at their final position, the surface tension causes filament thinning until the point where the filament breaks. On the other hand, for $\dot{\varepsilon}_0 > \dot{\varepsilon}_c$, the fluid solidifies at early stages within the stretching zone and the filament breaks before the top endplate reaches its final position by means of the combined action of pulling forces and surface tension. This behavior is seen in figure 9 (a and b) for the two larger particles, in all but the smallest strain rates with the value of $\dot{\varepsilon}_c$ decreasing with increasing particle size.

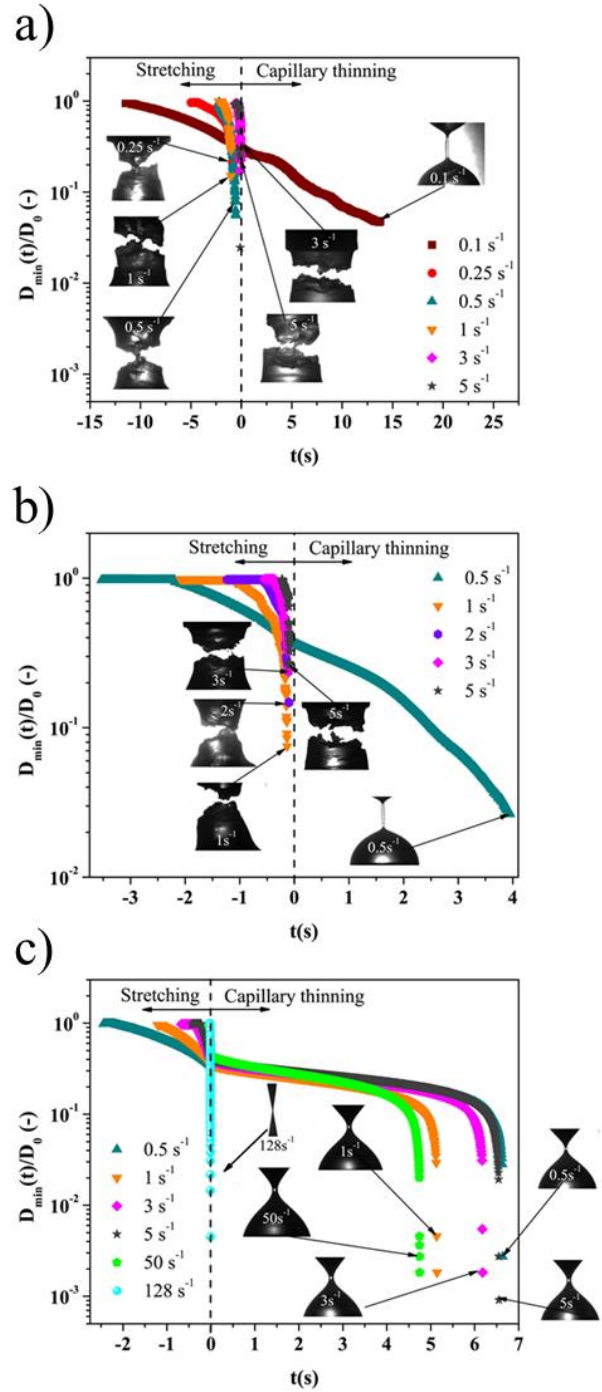

Figure 9. CaBER set-up experiments: Comparison between the time evolution of $D_{min}$ at different samples, and different initial strain rate rates, $\dot{\varepsilon}_0$, as indicated, for: (a) sample with particle radius R ≈ 689 nm, (b) for particles with R ≈ 405 nm and (c) for particles with R ≈ 137 nm. All samples were at a volume fraction of $\varphi$ ≈ 0.60.

For the two larger particle sizes (R ≈ 405 and R ≈ 689 nm) we see a filament stretching and breakup as a viscoelastic liquid only at the lowest strain rate measured (Figures 9a and 9b). This viscoelastic behavior at $\dot{\varepsilon}_0 < \dot{\varepsilon}_c$ is combined with a filament recoil at the end when the filament breaks. In Figure 9a and 9b we can observe a difference in filament dynamics for $\dot{\varepsilon}_0 < \dot{\varepsilon}_c$, between particles with R ≈ 405 and R ≈ 689 nm, with the filament thinning occurring faster in suspension with particles of R ≈ 405 nm. Actually for suspensions with the larger particles (R ≈ 689 nm), the filament seems to decelerate at the end, which is usually associated with an increase of the first normal stress difference in the filament as it was reported for liquid jets [81] and demonstrated recently for non-Brownian shear thickening fluids [82]. This is another demonstration of the of particle size dependence on the onset of jamming under extensional flow. For the smaller particle size (R ≈ 137 nm), the final filament breaks as in a Newtonian liquid for $\dot{\varepsilon}_0 < \dot{\varepsilon}_c$ (Figure 8e), and thinning dynamics is determined by the Newtonian behavior of the suspending fluid, as it has been reported before [79,83,84].

For, $\dot{\varepsilon}_0 > \dot{\varepsilon}_c$ samples with the two larger particles (R ≈ 405 nm and R ≈ 689 nm), exhibit a brittle failure reminiscent to crystalline solid materials. Under these conditions the samples fracture without any apparent yielding and plastic deformation observed (Figures 8b and 8d). Experiments with the smallest particles (R ≈ 137 nm) however reveal a highly ductile behavior under elongation, with the sample flowing until it necks, due to plastic deformation allowed in the microstructure (Figure 8f). Analogously to the failure mechanics of solid materials [85], there is a critical particle size, $R_c$, at which a 'nil-ductility' transition from ductile (particles smaller than $R_c$) to brittle failure will take place. In jammed suspensions of large particles, the reduced local mobility of the particles does not allow any microscopic rearrangements under elongation leading to brittle fracture at high stresses (or strain rates). On the other hand, similar stresses (or rates) induce a ductile failure in a sample with the smallest particles, as local particle mobility is higher (represented by shorter Brownian times). In fact, the ductile failure observed for the smallest particles (R ≈ 137 nm), is represented by filament shapes resembling those detected in a variety of yield stress fluids which exhibit similar break-up shapes, although they do not necessarily have the same microstructure [86–91]. On the other hand, at large strain rates $\dot{\varepsilon}_0 \gg \dot{\varepsilon}_c$ promote brittleness with a horizontal, but irregular fracture lines, (see Figure 8 and 9), similarly to what is expected in solid mechanics [85]. Smith *et al.* [59] also observed similar behavior for a system with similar PMMA particles in octadecene and suggested this could be the indication of the onset of jamming due to the self-filtration [92] of the system. This is caused at the microscopic level by a differential motion between the particles and the fluid, so that particles are left behind in the same position as the fluid drains during the filament thinning, as an effect of the combined force exerted on the sample by the moving plates and surface tension. This mechanism introduces heterogeneities in particle volume fraction during flow, which influences the dynamics of the filament during the thinning process.

*FSR experiments*

Filament stretching measurements were also performed by a FSR set-up with the aim to monitor the stress as a function of deformation in these HS colloidal glasses. For the smaller particle samples (R ≈ 137 nm), a liquid-like behavior was always seen in the range of the equipment (maximum strain rate 20 s$^{-1}$). For samples with particle size R ≈ 405 nm and R ≈ 689 nm we find again a critical strain rate, $\dot{\varepsilon}_c$ above which the filament structure changes and the sample shows a solid-like response with an irregular shape and rough texture due to dilatancy as shown in Figure 1b, similarly to CaBER experiments (Figure 8 and 9). However, due to the difficulty to perform measurements using the close loop control, since the filament shape is irregular, all measurements for $\dot{\varepsilon}_0 > \dot{\varepsilon}_c$ were performed without it. Therefore the equipment applied a constant strain rate controlling the top plate velocity [67], by checking the mid-filament diameter as a function of time. The critical strain rate $\dot{\varepsilon}_c$ is about 3.0 s$^{-1}$ (corresponding to $Pe$ ≈ 6.5) and 0.2 s$^{-1}$ ($Pe$ ≈ 4.5) for particles with R ≈ 405 nm and R ≈ 689 nm, respectively. Considering these values, it should be noted that the strain rate measured in the FSR is deduced using the filament mid-point diameter, while the strain rate in the CaBER is estimated from the velocity of the moving plates. Hence, we find that the critical strain rates $\dot{\varepsilon}_c$ in the CaBER is around 0.5 s$^{-1}$ ($Pe$ ≈ 2.3) and 0.1 s$^{-1}$ ($Pe$ ≈ 2.3) for the HS glasses with particle size R ≈ 405 nm and R ≈ 689 nm, respectively. The lower values of the critical rate in the CaBER can be justified not only by the difference in the method, but also by the fact that in FSR measurements the mid-filament plane was measured by a laser although jamming and necking be localized in neighboring plane (where the deformation/stress is different) while CaBER data were deduced at $D_{min}$ (and not at $D_{mid}$) from high speed camera images that provide more accurate profiling of the filament (see Figure 1a). Moreover, the stress and the critical rate may change if the initial sample size (aspect ratio) is varied; in this work, we kept the initial sample size the same.

Our main aim however in using the FSR was to obtain complementary information by measuring the stress during extensional flow and jamming of HS colloidal glasses. In Figure 10a (top) we show the measured force from a FSR set-up for strain rates below the critical rate for the observation of solid-like response as a function of Henky strain, for the three different particle size systems as indicated. For such low rates (i.e. for $\dot{\varepsilon}_0 < \dot{\varepsilon}_c$) the measured forces are rather small, and the data are quite scattered for all three samples measured; therefore no quantitate estimation of the stress during the extensional shear is in practice possible.

For high rates however, at $\dot{\varepsilon}_0 > \dot{\varepsilon}_c$ samples show the solid-like response seen with the CaBER set-up with the stretched filaments surface turning rough and opaque while the sample thins irregularly and eventually granulating and breaking as shown in figure 10b (top). In these cases, the force is much higher than at low strain rates, and clearly detectable by the instrument. Then the engineering stress, defined as $\sigma_{eng} = 4F(t)/(\pi D_0^2)$, can be determined and plotted (Figure 10b) as a function of Hencky strain. Although the quantitative reproducibility is weak, qualitatively different measurements show the same features. At this high strain rate regime, both samples shown in figure 10b show a stress increase with deformation and exhibit a stress peak at a certain Hencky strain. Correlating the stress response with simultaneous imaging of the sample we see that the stress increase, indicative of strain-hardening of the sample, takes place while the sample is still connected to the two plates while as the filament starts to exhibit necking and break the stress starts to decay till it drops to zero when the filament is fully broken.

As has been already reported [93] in weakly-strain-hardening fluids the engineering stress under extensional flow passes through a maximum when the elastic filament is unstable and starts necking, but grows without bounds at larger values of Deborah number (*De*). Analogous results have been published for shear thickening fluids by Rothstein and co-workers [57,58], which reported that just prior to the onset of filament failure, the extensional viscosity, which is proportional to the engineering stress measured in the FiSER, tends to go through a maximum. In the case of shear thickening fluids consisting of fumed silica particles with fractal chain-like structures, light scattering measurements showed that the extensional hardening was due to the alignment of nanoparticles and the formation of long strings of aggregates in the flow direction [58]. For cornstarch particles, which are nearly symmetric, the speed and magnitude of strain hardening increases remarkably fast with extension rate. At low extension rates, the fluid exhibits a Newtonian response while at moderate extension rates, it shows a weak strain hardening and forms long coherent fluid filaments. At a critical extension rate however, a dramatic increase in both the rate and magnitude of the strain hardening was observed with increasing extensional rate. Such dramatic increase in strain hardening was attributed to particle aggregation and the formation of an interconnected jammed network of clusters across the filament with a finite maximum strength independent of the extensional rate [57].

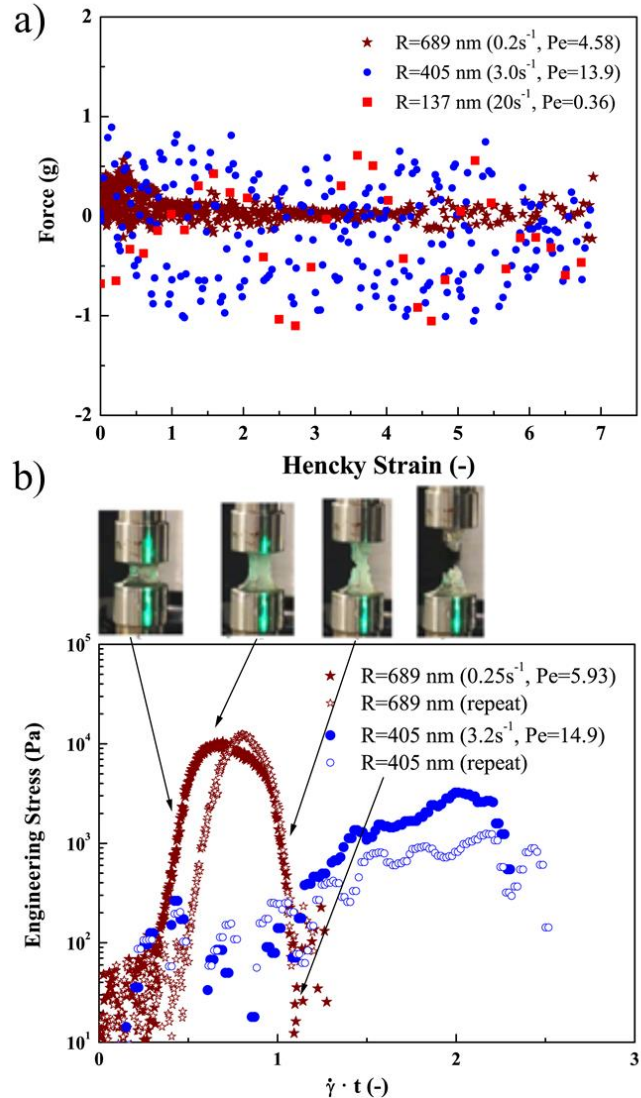

Figure 10. Experiments in a FSR extensional set-up for low (a) and high strain rates (b). The measured force versus the Hencky strain for all samples at $\dot{\varepsilon}_0 < \dot{\varepsilon}_c$ are shown in (a) similarly, in (b) we show the measured engineering stress versus the Hencky strain at $\dot{\varepsilon}_0 > \dot{\varepsilon}_c$ together with the corresponding images of the sample at different instances during stretching, as indicated by the arrows.

Here the samples with the larger particles (R ≈ 689 nm) seem to exhibit such behavior and develop a stress peak which is more pronounced and take place at lower strains compared to smaller particles (R ≈ 405 nm). These observations indicate the combined effect of the particle size and extension rate on the jamming phenomena due to the agglomeration of the particles and subsequent granulation of the sample suggesting that the mechanism responsible for jamming of concentrated HS suspensions and glasses during extensional flow is similar to that under shear. However, the critical deformation rates for developing such phenomena are lower in extensional than in shear flow. Our data show that for particles with R ≈ 405 nm and R ≈ 689 nm jamming occurs at $Pe ≈ 13$ during the shear flow and $Pe ≈ 2.3$ during extensional flow when measured by the CaBER, while when the FSR is used we get a critical shear rate of $Pe ≈ 6.5$ and $Pe ≈ 4.5$ for particles with R ≈ 405 nm and R ≈ 689 nm, respectively. Although there are differences between CaBER and FSR results, as discussed above, all measurements of the critical $Pe$ are clearly lower than that in shear experiments. The reason for this might be associated with confinement effects introduced in extensional flow due to a compressive flow in the radial direction that may increase packing, and subsequently promote local particle aggregation during axial stretching. It should be noted that at low initial aspect ratios ($\Lambda_0$), there is a contribution in the measured stress that arises from the shear components. As mentioned earlier, this is caused by the no-slip boundary condition at the end plates and is more prominent for samples with small aspect ratios. This effect could be accounted for using an empirical correction factor as suggested by Rasmussen et al. [94] for homogeneous polymer systems. However, we did not use the correction factor here since we clearly have an inhomogeneous system due to a dilatancy of the material. Therefore, the engineering stress presented in Figure 10b may vary for samples with different aspect ratios and may be influenced by the filament shape as well; thus should be used only qualitatively. Our findings clearly suggest that a more systematic investigation of the role of these variables in colloidal suspensions is needed.

In summary, the extensional behavior of HS glasses at a specific volume fraction depends on particle size and applied extension rate. We can infer that the conditions imposed in the CaBER set-up, namely the initial extension rate $\dot{\varepsilon}_0$, determine the arrangement of the particles during flow resulting in filament thinning processes with different shapes and breaking times. Figure 11 shows a master curve in a dimensionless diagram, where the time it takes for the filament to break up since the top plate start moving from its initial position at $h_0$, normalized with the Brownian time ($t_{br}/t_B$), is plotted against the Péclet number, $Pe = \dot{\varepsilon}_0 t_B$, for extensional experiments in the CaBER setup. This diagram suggests that samples show liquid-like behavior when the characteristic breaking time in the experiment is larger than the Brownian time ($t_{br}/t_B > 1$) otherwise, the behavior is solid-like. This is very reminiscent of arguments for the onset of shear thickening in hard-sphere colloids [28]; the onset stress for shear thickening, $\sigma_{min} = \eta \dot{\gamma}_{min}$, is related also with an onset shear rate (or $Pe$), reflecting the fact that an external shear stress must exceed all local stress barriers that prevent relative shear between particles. As long as the extensional deformation is performed slower than stress relaxation induced by local particle diffusion (related with the Brownian time) the sample exhibits a fluid like response. When this is not the case, the actual filament breakage time becomes smaller than the Brownian time and the sample starts behaving as a jammed granular system manifesting dilatancy and fracture.

The same observation also applies to the FSR experiments suggesting it is independent of the set-up used to induce extensional flow. More interestingly, the diagram indicates the distinction between Newtonian ($Pe < 1$) and non-Newtonian ($Pe > 1$) behaviors, within the liquid-like regime ($t_{br}/t_B < 1$) as well as between ductile ($Pe < 20$) and brittle response ($Pe > 20$), within the solid-like regime ($t_{br}/t_B > 1$). Table II summarizes this behavior and indicates that the critical Péclet number for the onset of granulation in the CaBER is $Pe_c \approx 2.3$, in the FSR set-up is on average $Pe_c \approx 5$, while in shear experiments $Pe_c \approx 13$ (Figure 4, where $N_1 > 0$), independently of the particle size. The fact of having lower critical $Pe$ values for the extensional flow than the steady shear is consistent with the flow dynamics associated with each type of experiment. In steady shear flow, the separation of two particles is linear, unlike in extensional flow, in which the separation is exponential as time increases [95].

For an ideal uniaxial extensional and shear experiments we expect that $Pe_{ext}/Pe_{shear} = \frac{1}{\sqrt{3}}$, since $\dot{\varepsilon}_0 = \frac{\dot{\gamma}}{\sqrt{3}}$ [95]. However, as stated above, during the early stages of both the CaBER and the FSR measurements besides the extension rate there is an additional contribution from a shear component proportional to $\frac{\dot{\varepsilon}_0}{\Lambda_0^2}$, i.e. $\dot{\gamma}^* = \frac{k\dot{\varepsilon}}{\Lambda_0^2}$. Thus, in the simplest case that $k = 1$ the strain rate tensor in the CaBER and the FSR is:

$$\bar{\bar{\dot{\gamma}}} = \begin{pmatrix} -\dot{\varepsilon}_0 & \frac{\dot{\varepsilon}_0}{\Lambda_0^2} & 0 \\ \frac{\dot{\varepsilon}_0}{\Lambda_0^2} & -\dot{\varepsilon}_0 & 0 \\ 0 & 0 & 2\dot{\varepsilon}_0 \end{pmatrix} \qquad (1)$$

with the magnitude of $\bar{\bar{\dot{\gamma}}}$ being:

$$\dot{\gamma}_{CaBER/FSR} = +\sqrt{\frac{1}{2}(\bar{\bar{\dot{\gamma}}}:\bar{\bar{\dot{\gamma}}})} = \dot{\varepsilon}_0 \sqrt{3 + \frac{1}{\Lambda_0^4}}. \qquad (2)$$

According to eqn. (2) and the experimental values of $\Lambda_0$ in the two set-ups, we should then expect that for the CaBER we get, $Pe_{c\_CaBER}/Pe_{c_{shear}} = \frac{1}{2.5} = 0.4$ and for the FSR, $Pe_{c_{FSR}}/Pe_{c_{shear}} = \frac{1}{6.8} = 0.147$, (instead of the ideal $Pe_{c\_ext}/Pe_{c\_shear} = \frac{1}{\sqrt{3}}$). In view of the above, the experimentally determined critical rates in the CaBER, $Pe_{c\_CaBER}/Pe_{c_{shear}} = 0.13$ is lower than the expected ratio (0.4), whereas for the FSR the experimental ratio, $Pe_{c\_FSR}/Pe_{c_{shear}} = 0.384$, is clearly higher than the expected value (0.147). This might be either due to a deviation of the true shear contribution present in the two experimental set-ups from the value estimated based on the $\Lambda_0$, term with $k = 1$ or to different microstructural mechanism of the HS glasses in shear and extension. With regard to the former, despite the smaller initial aspect ratio ($\Lambda_0$) in the FSR, the total shear contribution in the flow is apparently more than two times smaller than in the CaBER experiments and three times less than the expected value from eq. 2. The weaker shear contribution in the FSR may be attributed to the two-step stretching protocol used, where an initial stretch at extension rate well below the onset of extensional thickening indeed minimizes the shear effects on the second stretch, which is much faster and imposes a constant strain rate until the filament breaks up. This would quantify to a value of $k_{FSR} = 0.37$ assuming that the critical strain rate (for the onset of thickening and dilatation) is the same in extensional and shear experiments, hence attributing the differences in the critical strain rate values in shear and FSR experiments fully to an inaccurate determination of the shear contribution in the latter. A similar approach in the CaBER set-up would give $k_{CaBER} = 3.23$ since the shear contribution which is about three times higher than the expected according to eq. 2. We should note however that these $k$ should be taken cautiously as limiting values that indicate the amount of maximum shear contribution that could be present during the corresponding extensional experiments here.

Finally, the apparent extensional viscosity is also calculated from the CaBER experiments, based on equation $\eta_{Eapp} = -\frac{\sigma}{\frac{dDmin(t)}{dt}}$ [96], assuming the surface tension of the solvent as the surface tension of the sample. As it can be observed in Fgure 9, depending on the particle size and the initial extension rate imposed, the filament solidifies and breaks while the top plate is still travelling towards its final position ($h_f$). When the time to break is larger than the Brownian time, the sample does not solidify while being stretched from $h_0$ to $h_f$; this holds for *Pe* < 2.5 (Figure 11). Therefore, the apparent extensional viscosity can only be calculated once the top plate reaches the final position and the filament thins gradually until it breaks. The results are shown in Figure 12 as a function of the Péclet number. In all cases, the extensional viscosity is larger than 3 times the shear viscosity, and therefore the Trouton ratio, Tr, is larger than 3 even though granulation did not occur yet. This is mostly due the fact that we are observing a non-Newtonian behavior while Tr = 3 normally holds for Newtonian systems. Khandavalli and Rothstein [97] have also presented similar findings, which they attribute to non-spherical or fractal nature of the fumed silica particles used.

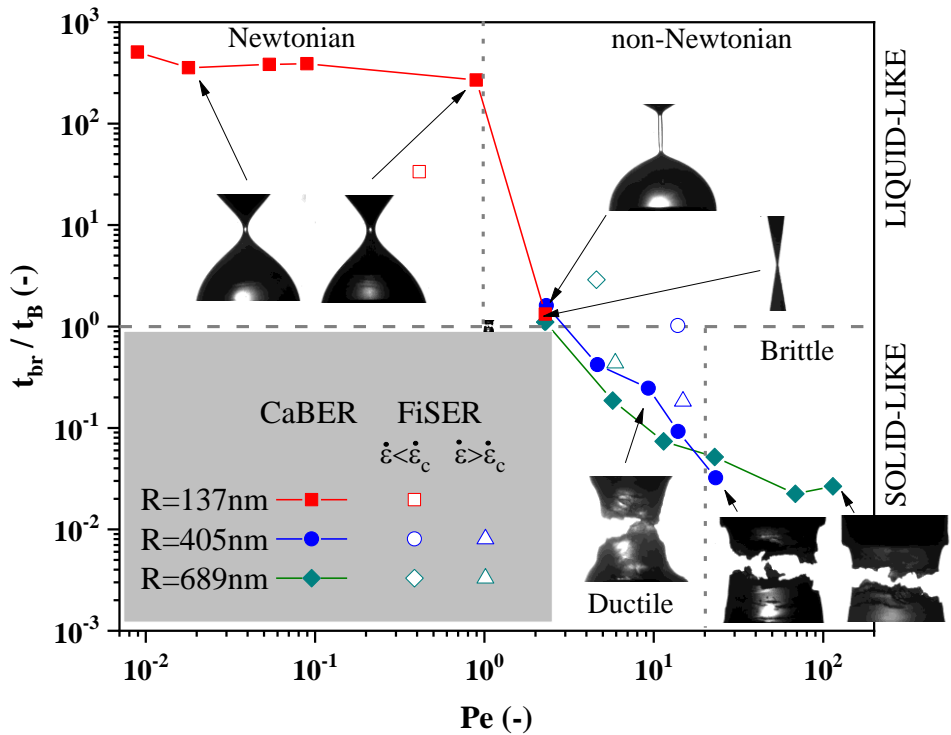

Figure 11. Dimensionless diagram of the different behavior shown by different samples at φ≈0.60 under elongational flow obtained with the CaBER set-up.

Table II. Summary of the different rheological behaviors observed in the dimensionless diagram for the samples at φ ≈ 0.60 under elongational flow as deduced CaBER experiments.

| $t_{br}/t_B$ | $Pe < 1$ | $1 < Pe < 2.5$ | $2.5 < Pe < 20$ | $Pe > 20$ |
|---|---|---|---|---|
| | | non-Newtonian | | |
| > 1 (liquid-like) | Newtonian | Viscoelastic-like Yield-Stress-like | | |
| <1 (solid-like) | | | Jamming Ductile | Jamming Brittle |

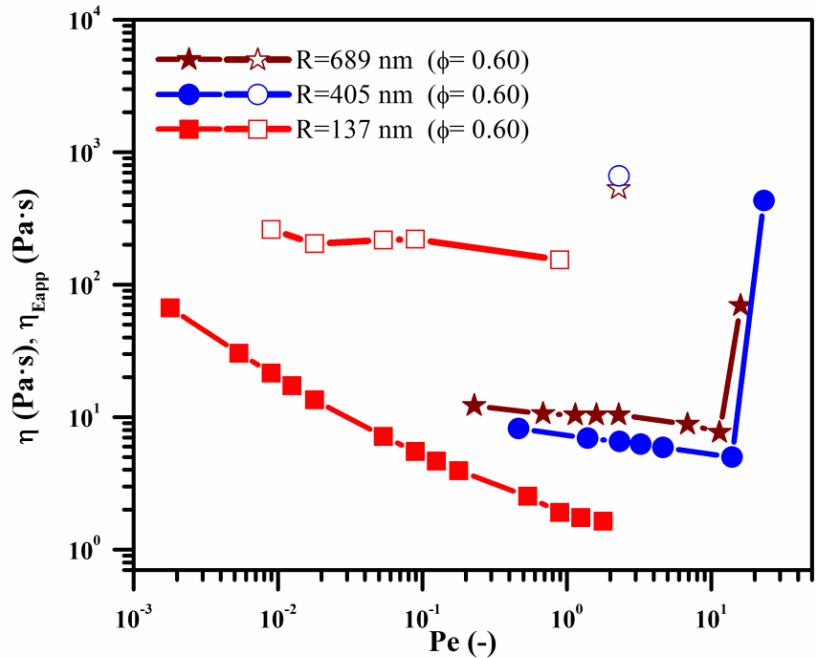

Figure 12. Comparison between the apparent shear (filled symbols) and extensional (empty symbols) viscosities of the samples at φ ≈ 0.6. Solid lines connect the experimental dots in order to guide the eye.

## 4. CONCLUSIONS

We have reported a dilatancy behavior of hard sphere colloidal glasses with different particle sizes under shear and extensional flow. Discontinuous shear thickening is observed for larger particles at lower critical shear rate values. Start-up shear experiments at different shear rates presented stress overshoots above the critical shear rate for the onset of DST, suggesting a change in the suspension structure during the flow. Moreover, $N_1$ measurements during step rate, have shown a transition from $N_1 \approx 0$ (at low shear rates), to $N_1 < 1$ (in the vicinity of critical shear rate) indicating contributions from lubrications forces, and finally $N_1 > 1$ (beyond critical shear rate) suggesting frictional interactions from contacts between particles. The critical Péclet number for the dramatic granulation in shear experiments (at $Pe_{cr} \approx 15$), is independent of the particle size. Moreover, sheared samples were monitored by video camera, revealing that dilatancy occurs at the presence of contact interactions ($N_1 > 0$) in all samples showing DST, i.e. systems with large particles (R ≈ 405 nm and R ≈ 689 nm). While dilatancy occurs for particles with R ≈ 689 nm at lower shear rates, the same phenomenon is detected for particles with R ≈ 405 nm, but at higher shear rates. This suggests that contact between particles depends on both, shear rate applied and size, but also could be, as it was recently stated, that hydrodynamics interactions could mask positive friction depending on particles size and flow conditions.

Dilatancy for suspensions with particle size R ≈ 405 nm and R ≈ 689 nm was also observed under extensional flow, using both CaBER and FSR set-ups, where a critical value of the strain rate ($\dot{\varepsilon}_c$) was detected indicating structural changes in the HS glass. Below a critical strain rate, liquid-like behavior was seen for all particle sizes, with viscoelastic behavior at the filament break for the large particle (R ≈ 405 nm and R ≈ 689 nm) samples, mostly due to self-filtration effects induced by filament thinning due to surface tension. This mechanism leads to heterogeneities of the particle volume fraction within the sample during flow and is related with differences in filament dynamics in the two different particle size (R ≈ 405 nm and R ≈ 689 nm) samples. The samples with the larger particles (R ≈ 689 nm), exhibits a slowing down of filament thinning, indicating the onset of creation of a jamming network under extensional flow, while for the samples with the smaller particles (R ≈ 137 nm) the filament thinning dynamics is Newtonian-like for $\dot{\varepsilon}_0 < \dot{\varepsilon}_c$. At high strain rates, for $\dot{\varepsilon}_0 > \dot{\varepsilon}_c$, large particle (R ≈ 405 nm and R ≈ 689 nm) samples exhibit a combination of granulation and brittle fracture. For smaller particles (R ≈ 137 nm) extensional flow at $\dot{\varepsilon}_0 > \dot{\varepsilon}_c$ leads to filament stretching similar to that observed in yield stress fluids which finally break with a ductile type fracture.

These findings were verified by the FSR measurements that similarly show liquid-like behavior for $\dot{\varepsilon}_0 < \dot{\varepsilon}_c$, and solid-like behavior when $\dot{\varepsilon}_0 > \dot{\varepsilon}_c$, both for suspensions with R ≈ 405 nm and R ≈ 689 nm particles. Here, jamming, dilatancy, and subsequent granulation were observed in both suspensions, within the range of the equipment (maximum strain rate 20 s$^{-1}$). Direct observation of stress measurements, showed the effect of particle size on jamming and granulation, suggesting that the mechanism governing this behavior under shear is also present in extensional flow. The strain rates involved in these phenomena are lower in extensional flow than in shear flow, mostly due to the confinement effect of the compressive flow in the radial direction that can increase packing, and subsequent the local aggregation of particles during the stretching.

Finally, the Péclet number scales the influence of flow on granulation, for samples with different particle sizes. In shear experiments, this phenomenon was observed above a critical value of $Pe_{cr} \approx 15$, while in extensional flow, in the case of CaBER at $Pe_{cr} \approx 2.5$, and in FSR at $Pe_{cr} \approx 5$. These findings are consistent with the characteristics of each type of flow: the separation of two particles is linear during simple shear flow, unlike in extensional devices where there is an unavoidably combination of extensional and shear flows at the early stage of the experiments, which is larger in the CaBER than in the FSR, and results in a much faster structural changes leading to shear thickening.

## 5. SUPPLEMENTARY MATERIAL

See the supplementary material to observe the results with cone plate and cone-partition-plate (CPP) in steady shear measurements as well as in oscillatory shear.

## 6. ACKNOWLEDGEMENT

The authors would like to express their sincere gratitude to Prof. João Maia for valuable and constructive suggestions in this work. Financial support from the EU Project "SmartPro" and National Project ARiSTEIA II "MicroSoft", the Aage og Johanne Louis-Hansen Foundation, Coordenação de Aperfeiçoamento de Pessoal de Nível Superior - Brasil (CAPES) - PRINT 88887.310339/2018-00, Fundação de Amparo a Pesquisa de São Paulo (FAPESP) for the grants 2012/50259-8, and Fundo Mackenzie de Pesquisa (MackPesquisa, Project Number 181009), is gratefully acknowledged. FJGR and LCD would also like to acknowledge financial support from Fundação para a Ciência e a Tecnologia (FCT), COMPETE and FEDER through grants IF/00190/2013 and IF/00148/2013, and project IF/00190/2013/CP1160/CT0003.

**References**

[1]   Barnes HA. Shear Thickening ("Dilatancy") in Suspensions of Nonaggregating Solid Particles Dispersed in Newtonian Liquids. J Rheol 1989;33:329–66. https://doi.org/doi:http://dx.doi.org/10.1122/1.550017.

[2]   Wagner NJ, Brady JF. Shear thickening in colloidal dispersions. Phys Today 2009;62:27–32. https://doi.org/10.1063/1.3248476.

[3]   Lee YS, Wetzel ED, Wagner NJ. The ballistic impact characteristics of Kevlar® woven fabrics impregnated with a colloidal shear thickening fluid. J Mater Sci n.d.;38:2825–33. https://doi.org/10.1023/a:1024424200221.

[4]   Galindo-Rosales FJ, Martínez-Aranda S, Campo-Deaño L. CorkSTFμfluidics – A novel concept for the development of eco-friendly light-weight energy absorbing composites. Mater Des 2015;82:326–34. https://doi.org/10.1016/j.matdes.2014.12.025.

[5]   Helber R, Doncker F, Bung R. Vibration attenuation by passive stiffness switching mounts. J Sound Vib 1990;138:47–57. https://doi.org/http://dx.doi.org/10.1016/0022-460X(90)90703-3.

[6]   Laun HM, Bung R, Schmidt F. Rheology of extremely shear thickening polymer dispersionsa) (passively viscosity switching fluids). J Rheol (N Y N Y) 1991;35:999–1034. https://doi.org/doi:http://dx.doi.org/10.1122/1.550257.

[7]   Nilsson MA, Kulkarni R, Gerberich L, Hammond R, Singh R, Baumhoff E, et al. Effect of fluid rheology on enhanced oil recovery in a microfluidic sandstone device. J Nonnewton Fluid Mech 2013;202:112–9.

[8]   Hoffman RL. Discontinuous and Dilatant Viscosity Behavior in Concentrated Suspensions. I. Observation of a Flow Instability. Trans Soc Rheol 1972;16:155–73. https://doi.org/doi:http://dx.doi.org/10.1122/1.549250.

[9]   Hoffman RL. Discontinuous and dilatant viscosity behavior in concentrated suspensions.

II. Theory and experimental tests. J Colloid Interface Sci 1974;46:491–506.

[10] Hoffman RL. Discontinuous and dilatant viscosity behavior in concentrated suspensions III. Necessary conditions for their occurrence in viscometric flows. Adv Colloid Interface Sci 1982;17:161–84.

[11] Egres RG, Wagner NJ. The rheology and microstructure of acicular precipitated calcium carbonate colloidal suspensions through the shear thickening transition. J Rheol 2005;49:719–46. https://doi.org/doi:http://dx.doi.org/10.1122/1.1895800.

[12] Egres RG, Nettesheim F, Wagner NJ. Rheo-SANS investigation of acicular-precipitated calcium carbonate colloidal suspensions through the shear thickening transition. J Rheol 2006;50:685–709. https://doi.org/doi:http://dx.doi.org/10.1122/1.2213245.

[13] Brady JF, Bossis G. The rheology of concentrated suspensions of spheres in simple shear flow by numerical simulation. J Fluid Mech 1985;155:105–29. https://doi.org/doi:10.1017/S0022112085001732.

[14] D'Haene P, Mewis J, Fuller GG. Scattering Dichroism Measurements of Flow-Induced Structure of a Shear Thickening Suspension. J Colloid Interface Sci 1993;156:350–8.

[15] Bender JW, Wagner NJ. Optical Measurement of the Contributions of Colloidal Forces to the Rheology of Concentrated Suspensions. J Colloid Interface Sci 1995;172:171–84.

[16] Bender J, Wagner NJ. Reversible shear thickening in monodisperse and bidisperse colloidal dispersions. J Rheol 1996;40:899–916. https://doi.org/doi:http://dx.doi.org/10.1122/1.550767.

[17] Laun HM, Bung R. Rheological and small angle neutron scattering investigation of shear-induced particle structures of concentrated polymer dispersions submitted to plane Poiseuille and Couette flow a). J Rheol 1992:743–87.

[18] Maranzano BJ, Wagner NJ. Flow-small angle neutron scattering measurements of colloidal dispersion microstructure evolution through the shear thickening transition. J Chem Phys 2002;117:10291–302. https://doi.org/doi:http://dx.doi.org/10.1063/1.1519253.

[19] Lee YS, Wagner NJ. Rheological Properties and Small-Angle Neutron Scattering of a Shear Thickening, Nanoparticle Dispersion at High Shear Rates. Ind Eng Chem Res 2006;45:7015–24.

[20] Kalman DP, Merrill RL, Wagner NJ, Wetzel ED. Effect of Particle Hardness on the Penetration Behavior of Fabrics Intercalated with Dry Particles and Concentrated

Particleâˆ'Fluid Suspensions. ACS Appl Mater Interfaces 2009;1:2602–12.

[21] Cheng X, McCoy JH, Israelachvili JN, Cohen I. Imaging the Microscopic Structure of Shear Thinning and Thickening Colloidal Suspensions. Science (80- ) 2011;333:1276–9. https://doi.org/10.1126/science.1207032.

[22] Mewis and N. J. Wagner J. Colloidal Suspension Rheology . 2012.

[23] Boersma WH, Laven J, Stein HN. Shear thickening(dilatancy) in concentrated dispersions. AIChE J 1990;36:321–32. https://doi.org/10.1002/aic.690360302.

[24] Bergenholtz J, Brady JF, Vicic M. The non-Newtonian rheology of dilute colloidal suspensions 2002.

[25] Fernandez N, Mani R, Rinaldi D, Kadau D, Mosquet M, Cayer-barrioz J, et al. Microscopic Mechanism for Shear Thickening of Non-Brownian Suspensions 2013;108301:1–5. https://doi.org/10.1103/PhysRevLett.111.108301.

[26] Heussinger C. Shear thickening in granular suspensions: Interparticle friction and dynamically correlated clusters. Phys Rev E 2013;88:050201. https://doi.org/10.1103/PhysRevE.88.050201.

[27] Seto R, Mari R, Morris JF, Denn MM. Discontinuous Shear Thickening of Frictional Hard-Sphere Suspensions. Phys Rev Lett 2013;111. https://doi.org/10.1103/PhysRevLett.111.218301.

[28] Brown M. Jaeger E and H. Shear thickening in concentrated suspensions: phenomenology, mechanisms and relations to jamming. Rep Prog Phys 2014;77:46602.

[29] Brown E, Jaeger HM. The role of dilation and confining stresses in shear thickening of dense suspensions. J Rheol 2012;56:875–923. https://doi.org/doi:http://dx.doi.org/10.1122/1.4709423.

[30] Fall A, Huang N, Bertrand F, Ovarlez G, Bonn D. Shear Thickening of Cornstarch Suspensions as a Reentrant Jamming Transition. Phys Rev Lett 2008;100:18301.

[31] Reynolds O. LVII. On the dilatancy of media composed of rigid particles in contact. With experimental illustrations. Philos Mag Ser 5 1885;20:469–81.

[32] Metzner AB, Whitlock M. Flow Behavior of Concentrated (Dilatant) Suspensions. Trans Soc Rheol 1958;2:239–54. https://doi.org/doi:http://dx.doi.org/10.1122/1.548831.

[33] Lootens D, van Damme H, Hémar Y, Hébraud P. Dilatant Flow of Concentrated Suspensions of Rough Particles. Phys Rev Lett 2005;95:268302.

[34]	Brown E, Jaeger HM. Dynamic Jamming Point for Shear Thickening Suspensions. Phys Rev Lett 2009;103:86001.

[35]	Holmes M. Fuchs and M. E. Cates CB. Jamming transitions in a schematic model of suspension rheology. Eur Lett 2003;63:240–246.

[36]	Melrose JR, Ball RC. "Contact networks" in continuously shear thickening colloids. J Rheol 2004;48:961–78. https://doi.org/doi:http://dx.doi.org/10.1122/1.1784784.

[37]	Wyart M, Cates ME. Discontinuous shear thickening without inertia in dense non-brownian suspensions. Phys Rev Lett 2014;112:1–5. https://doi.org/10.1103/PhysRevLett.112.098302.

[38]	Lin NYC, Guy BM, Hermes M, Ness C, Sun J, Poon WCK, et al. Hydrodynamic and Contact Contributions to Continuous Shear Thickening in Colloidal Suspensions. Phys Rev Lett 2015;115:228304. https://doi.org/10.1103/PhysRevLett.115.228304.

[39]	Guy BE, Hermes M, Poon WEE. Towards a Unified Description of the Rheology of Hard-Particle Suspensions. Phys Rev Lett 2015;115:1–5. https://doi.org/10.1103/PhysRevLett.115.088304.

[40]	Hsiao LC, Jamali S, Beltran-Villegas DJ, Glynos E, Green PF, Larson RG, et al. A rheological state diagram for rough colloids in shear flow 2016.

[41]	Jamali S, Boromand A, Wagner N, Maia J. Microstructure and rheology of soft to rigid shear-thickening colloidal suspensions. J Rheol (N Y N Y) 2015;59:1377–95. https://doi.org/10.1122/1.4931655.

[42]	Phung TN, Brady JF, Bossis G. Stokesian Dynamics simulation of Brownian suspensions. J Fluid Mech 1996;313:181–207. https://doi.org/doi:10.1017/S0022112096002170.

[43]	FOSS DR, BRADY JF. Structure, diffusion and rheology of Brownian suspensions by Stokesian Dynamics simulation. J Fluid Mech 2000;407:S0022112099007557. https://doi.org/10.1017/S0022112099007557.

[44]	Melrose JR, Ball RC. Continuous shear thickening transitions in model concentrated colloids—The role of interparticle forces. J Rheol (N Y N Y) 2004;48:937. https://doi.org/10.1122/1.1784783.

[45]	Brown E, Jaeger HM. The role of dilation and confining stresses in shear thickening of dense suspensions. J Rheol (N Y N Y) 2012;56:875. https://doi.org/10.1122/1.4709423.

[46]	R. Byron Bird, Robert C. Armstrong OH. Dynamics of Polymeric Liquids. Volume 1.

Fluid mechanics. JOHN WILEY & Sons; 1987.

[47]  Macosko CW. Rheology : principles, measurements, and applications. VCH; 1994.

[48]  Cates ME, Wittmer JP, Bouchaud J-P, Claudin P. Jamming, Force Chains, and Fragile Matter. Phys Rev Lett 1998;81:1841–4. https://doi.org/10.1103/PhysRevLett.81.1841.

[49]  Cates ME, Haw MD, Holmes CB. Dilatancy, jamming, and the physics of granulation. J Phys Condens Matter 2005;17:S2517–31. https://doi.org/10.1088/0953-8984/17/24/010.

[50]  Gurnon AK, Wagner NJ, Ackerson BJ, Hayter JB, Clark NA, Cotter L, et al. Microstructure and rheology relationships for shear thickening colloidal dispersions. J Fluid Mech 2015;769:242–76. https://doi.org/10.1017/jfm.2015.128.

[51]  Cwalina CD, Wagner NJ. Material properties of the shear-thickened state in concentrated near hard-sphere colloidal dispersions. J Rheol (N Y N Y) 2014;58:949–67. https://doi.org/10.1122/1.4876935.

[52]  Laun HM. Normal stresses in extremely shear thickening polymer dispersions. J Nonnewton Fluid Mech 1994;54:87–108. https://doi.org/10.1016/0377-0257(94)80016-2.

[53]  Lee M, Alcoutlabi M, Magda JJ, Dibble C, Solomon MJ, Shi X, et al. The effect of the shear-thickening transition of model colloidal spheres on the sign of N[sub 1] and on the radial pressure profile in torsional shear flows. J Rheol (N Y N Y) 2006;50:293. https://doi.org/10.1122/1.2188567.

[54]  Royer JR, Blair DL, Hudson SD. Rheological Signature of Frictional Interactions in Shear Thickening Suspensions 2016;188301:1–5. https://doi.org/10.1103/PhysRevLett.116.188301.

[55]  Sankaran AK, Rothstein JP. Effect of viscoelasticity on liquid transfer during gravure printing. J Nonnewton Fluid Mech 2012;175:64–75.

[56]  Galindo-Rosales FJ, Alves MA, Oliveira MSN. Microdevices for extensional rheometry of low viscosity elastic liquids: a review. Microfluid Nanofluidics 2013;14:1–19. https://doi.org/10.1007/s10404-012-1028-1.

[57]  Bischoff White E, Chellamuthu M, Rothstein J. Extensional rheology of a shear-thickening cornstarch and water suspension. Rheol Acta 2010;49:119–29.

[58]  Chellamuthu M, Arndt EM, Rothstein JP. Extensional rheology of shear-thickening nanoparticle suspensions. Soft Matter 2009;5:2117–24.

[59]  Smith MI, Besseling R, Cates ME, Bertola V. Dilatancy in the flow and fracture of

stretched colloidal suspensions. Nat Commun 2010;1:114.

[60] Royall CP, Poon WCK, Weeks ER. In search of colloidal hard spheres. Soft Matter 2013;9:17–27. https://doi.org/10.1039/C2SM26245B.

[61] Maranzano BJ, Wagner NJ. The effects of particle size on reversible shear thickening of concentrated colloidal dispersions. J Chem Phys 2001;114:10514–27. https://doi.org/doi:http://dx.doi.org/10.1063/1.1373687.

[62] Tuladhar TR, Mackley MR. Filament stretching rheometry and break-up behaviour of low viscosity polymer solutions and inkjet fluids. J Nonnewton Fluid Mech 2008;148:97–108. https://doi.org/10.1016/j.jnnfm.2007.04.015.

[63] Rodd LE, Scott TP, Cooper-White JJ, McKinley GH. Capillary break-up rheometry of low-viscosity elastic fluids. Appl Rheol 2005;15:12–27. https://doi.org/10.3933/ApplRheol-15-12.

[64] Campo-Deaño L, Clasen C. The slow retraction method (SRM) for the determination of ultra-short relaxation times in capillary breakup extensional rheometry experiments. J Nonnewton Fluid Mech 2010;165:1688–99. https://doi.org/http://dx.doi.org/10.1016/j.jnnfm.2010.09.007.

[65] Galindo-Rosales FJ, Segovia-Gutiérrez JP, Pinho FT, Alves MA, de Vicente J. Extensional rheometry of magnetic dispersions. J Rheol (N Y N Y) 2015;59:193–209. https://doi.org/10.1122/1.4902356.

[66] Niedzwiedz K, Arnolds O, Willenbacher N, Brummer R. How to characterize yield stress fluids with capillary breakup extensional rheometry (CaBER)? Appl Rheol 2009;19:1–10. https://doi.org/10.3933/ApplRheol-19-41969.

[67] Spiegelberg SH, Ables DC, McKinley GH. The role of end-effects on measurements of extensional viscosity in filament stretching rheometers. J Nonnewton Fluid Mech 1996;64:229–67. https://doi.org/10.1016/0377-0257(96)01439-5.

[68] McKinley GH, Sridhar T. F ILAMENT -S TRETCHING R HEOMETRY OF C OMPLEX F LUIDS. Annu Rev Fluid Mech 2002;34:375–415. https://doi.org/10.1146/annurev.fluid.34.083001.125207.

[69] Bach A, Rasmussen HK, Hassager O. Extensional viscosity for polymer melts measured in the filament stretching rheometer. J Rheol (N Y N Y) 2003;47:429–41. https://doi.org/10.1122/1.1545072.

[70] Román Marín JM, Huusom JK, Alvarez NJ, Huang Q, Rasmussen HK, Bach A, et al. A control scheme for filament stretching rheometers with application to polymer melts. J Nonnewton Fluid Mech 2013;194:14–22. https://doi.org/10.1016/j.jnnfm.2012.10.007.

[71] Orr N V., Sridhar T. Probing the dynamics of polymer solutions in extensional flow using step strain rate experiments. J Nonnewton Fluid Mech 1999;82:203–32. https://doi.org/10.1016/S0377-0257(98)00163-3.

[72] Maranzano BJ, Wagner NJ. The effects of interparticle interactions and particle size on reversible shear thickening: Hard-sphere colloidal dispersions. J Rheol 2001;45:1205–22. https://doi.org/doi:http://dx.doi.org/10.1122/1.1392295.

[73] Koumakis N, Laurati M, Egelhaaf SU, Brady JF, Petekidis G. Yielding of Hard-Sphere Glasses during Start-Up Shear. Phys Rev Lett 2012;108:98303.

[74] Koumakis N, Laurati M, Jacob AR, Mutch KJ, Abdellali A, Schofield AB, et al. Start-up shear of concentrated colloidal hard spheres: Stresses, dynamics, and structure. J Rheol (N Y N Y) 2016;60:603–23. https://doi.org/10.1122/1.4949340.

[75] Snijkers F, Vlassopoulos D. Cone-partitioned-plate geometry for the ARES rheometer with temperature control. J Rheol (N Y N Y) 2011;55:1167–86. https://doi.org/10.1122/1.3625559.

[76] Koumakis N, Laurati M, Jacob AR, Mutch KJ, Abdellali A, Schofield AB, et al. Start-up shear of concentrated colloidal hard spheres: Stresses, dynamics, and structure. J Rheol (N Y N Y) 2016;60:603–23. https://doi.org/10.1122/1.4949340.

[77] Hermes M, Guy BM, Poy G, Cates ME, Wyart M, Poon WCK. Unsteady flow and particle migration in dense, non-Brownian suspensions. ArXiv 2015. https://doi.org/10.1122/1.4953814.

[78] Jacob AR, Moghimi E, Petekidis G. Rheological signatures of aging in hard sphere colloidal glasses. Phys Fluids 2019;31:087103. https://doi.org/10.1063/1.5113500.

[79] Mathues W, McIlroy C, Harlen OG, Clasen C. Capillary breakup of suspensions near pinch-off. Phys Fluids 2015;27:093301. https://doi.org/10.1063/1.4930011.

[80] Sadek SH, Najafabadi HH, Galindo-Rosales FJ. Capillary breakup extensional magnetorheometry. J Rheol (N Y N Y) 2020;64:55–65. https://doi.org/10.1122/1.5115460.

[81] Eggers J, Villermaux E. Physics of liquid jets. Rep Prog Phys 2008;71:36601–79. https://doi.org/10.1088/0034-4885/71/3/036601.

[82] Roché M, Kellay H, Stone HA. Heterogeneity and the role of normal stresses during the extensional thinning of non-Brownian shear-thickening fluids. Phys Rev Lett 2011;107:134503. https://doi.org/10.1103/PhysRevLett.107.134503.

[83] Furbank RJ, Morris JF. An experimental study of particle effects on drop formation. Phys Fluids 2004;16:1777. https://doi.org/10.1063/1.1691034.

[84] Furbank RJ, Morris JF. Pendant drop thread dynamics of particle-laden liquids. Int J Multiph Flow 2007;33:448–68. https://doi.org/10.1016/j.ijmultiphaseflow.2006.02.021.

[85] Roylance D. MECHANICAL PROPERTIES OF MATERIALS 2008.

[86] Niedzwiedz K Willenbacher N, Brummer R AO. Capillary Breakup Extensional Rheometry of Yield Stress Fluids. Appl Rheol 19 2009:41969.

[87] Herschel W. H. and Bulkley R. Measurement of Consistency as Applied to Rubber-Benzene Solutions. Proc, Am Soc Test Mater 1926;26:621.

[88] Q D Nguyen and, Boger D V. Measuring the Flow Properties of Yield Stress Fluids 2003.

[89] Barnes HA. The yield stress—a review or 'παντα ρει'—everything flows? J Nonnewton Fluid Mech 1999;81:133–78. https://doi.org/10.1016/S0377-0257(98)00094-9.

[90] Bonn D, Denn MM. Materials science. Yield stress fluids slowly yield to analysis. Science 2009;324:1401–2. https://doi.org/10.1126/science.1174217.

[91] Ovarlez G, Cohen-Addad S, Krishan K, Goyon J, Coussot P. On the existence of a simple yield stress fluid behavior. J Nonnewton Fluid Mech 2013;193:68–79. https://doi.org/10.1016/j.jnnfm.2012.06.009.

[92] Haw MD. Jamming, Two-Fluid Behavior, and "Self-Filtration" in Concentrated Particulate Suspensions. Phys Rev Lett 2004;92:185506. https://doi.org/10.1103/PhysRevLett.92.185506.

[93] Minwu Yaoa, Stephen H. Spiegelbergb GHM. Dynamics of weakly strain-hardening fluids in filament stretching devices. J Nonnewton Fluid Mech 2000;89:1–43. https://doi.org/10.1016/S0377-0257(99)00028-2.

[94] Rasmussen HK, Bejenariu AG, Hassager O, Auhl D. Experimental evaluation of the pure configurational stress assumption in the flow dynamics of entangled polymer melts. J Rheol (N Y N Y) 2010;54:1325. https://doi.org/10.1122/1.3496378.

[95] Morrison FA. Understanding rheology. Oxford University Press; 2001.

[96] Anna SL, McKinley GH. Elasto-capillary thinning and breakup of model elastic liquids. J

Rheol (N Y N Y) 2001;45:115. https://doi.org/10.1122/1.1332389.

[97]  Khandavalli S, Rothstein JP. Extensional rheology of shear-thickening fumed silica nanoparticles dispersed in an aqueous polyethylene oxide solution. J Rheol (N Y N Y) 2014;58:411–31. https://doi.org/10.1122/1.4864620.